\documentclass[aps]{revtex4}
\usepackage{graphicx}
\usepackage{amsmath}
\usepackage{amsfonts}

\begin{document}
\title{Phase-matching-free parametric oscillators based on two dimensional semiconductors}
\author{A. Ciattoni$^{1}$}
\email{alessandro.ciattoni@spin.cnr.it}
\author{A. Marini$^{2}$}
\email{andrea.marini@icfo.es}
\author{C. Rizza$^{3}$}
\author{C. Conti$^{4,5}$}
\affiliation{$^1$Consiglio Nazionale delle Ricerche, CNR-SPIN, Via Vetoio 10, 67100 L'Aquila, Italy}
\affiliation{$^2$ICFO-Institut de Ciencies Fotoniques, The Barcelona Institute of Science and Technology, 08860 Castelldefels (Barcelona), Spain}
\affiliation{$^3$Department of Industrial and Information Engineering and Economics, Via G. Gronchi 18, University of L'Aquila, I-67100 L'Aquila, Italy}
\affiliation{$^4$Institute for Complex Systems (ISC-CNR), Via dei Taurini 19, 00185, Rome, Italy}
\affiliation{$^5$Department of Physics, University Sapienza, Piazzale Aldo Moro 5, 00185, Rome, Italy}
\date{\today}

\begin{abstract}
Optical parametric oscillators are widely-used pulsed and continuous-wave tunable sources for innumerable applications, as in quantum technologies, imaging and biophysics. A key drawback is material dispersion imposing the phase-matching condition that generally entails a complex setup design, thus hindering tunability and miniaturization. Here we show that the burden of phase-matching is surprisingly absent in parametric micro-resonators adopting monolayer transition-metal dichalcogenides as quadratic nonlinear materials. By the exact solution of nonlinear Maxwell equations and first-principle calculation of the semiconductor nonlinear response, we devise a novel kind of phase-matching-free miniaturized parametric oscillator operating at conventional pump intensities. We find that different two-dimensional semiconductors yield degenerate and non-degenerate emission at various spectral regions thanks to  doubly-resonant mode excitation, which can be tuned through the incidence angle of the external pump laser. In addition we show that high-frequency electrical modulation can be achieved by doping through electrical gating that efficiently shifts the parametric oscillation threshold. Our results pave the way for new ultra-fast tunable micron-sized sources of entangled photons, a key device underpinning any quantum protocol. Highly-miniaturized optical parametric oscillators may also be employed in lab-on-chip technologies for biophysics, environmental pollution detection and security.
\end{abstract}

\maketitle

\section{Introduction}
Optical nonlinearity in photonic materials enables an enormous amount of applications such as frequency conversion \cite{Franken1961,Birks2000}, all-optical signal processing \cite{Stegeman1996,Koos2009}, and non-classical sources \cite{Kwiat1995,Stevenson2006}. Parametric down-conversion (PDC) furnishes tunable sources of coherent radiation \cite{Giordmaine1965,Yariv1966,Brosnan1979,Eckardt1991,Debuisschert1993,Fabre1997,Levy2010,Razzari2010} and generators of entangled photons and squeezed states of light \cite{Wu1987,Lu2000}. In traditional configurations, a nonlinear crystal with broken centrosymmetry and second-order nonlinearity sustains PDC \cite{Giordmaine1965,Yariv1966,Brosnan1979,Eckardt1991,Debuisschert1993,Fabre1997}; more recently, effective PDC was reported in centrosymmetric crystals with third-order nonlinearity \cite{Levy2010,Razzari2010} and semiconductor microcavities \cite{Ciuti2003,Diederichs2006,Abbarchi2011}.
\begin{figure*}
\includegraphics[width=1\textwidth]{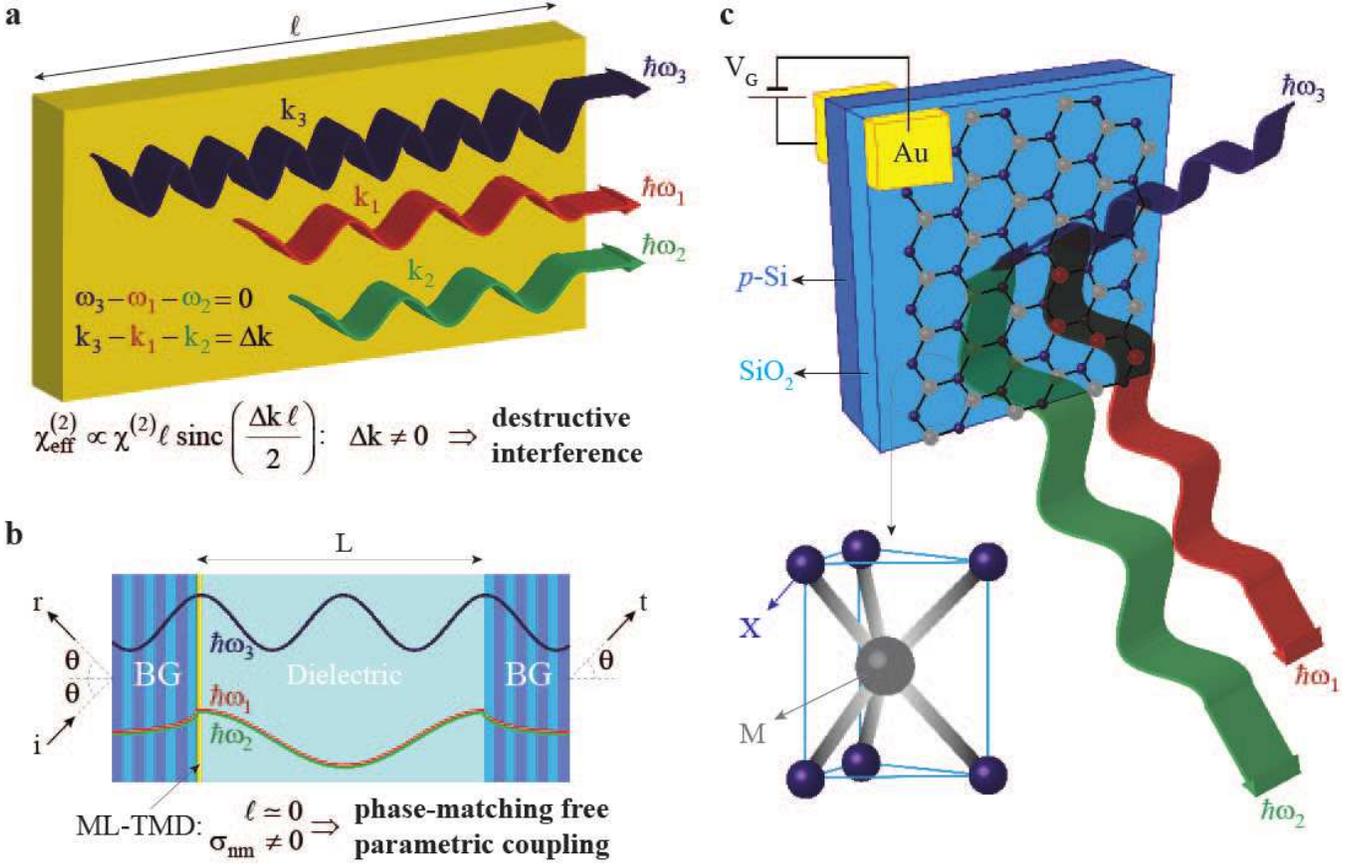}
\caption{{\bf Phase-matching free micron-sized parametric oscillators.} {\bf a.} Schematic illustration of conventional three-wave parametric coupling in bulk nonlinear crystals. The effective quadratic susceptibility $\chi_{eff}^{(2)}$ is heavily affected by the mismatch $\Delta k$ among the wavevectors $k_m = n_m \omega_m /c$ of the pump (3), signal (1) and idler (2) waves whose destructive interference $\Delta k \neq 0 $ hinders parametric coupling. {\bf b.} Sketch of the ML-TMD based parametric oscillator. The cavity is assembled by two Bragg mirrors separated by a dielectric layer and the ML-TMD is placed on the left mirror. The incident (i) pump field produces both reflected (r) and transmitted (t) pump, signal and idler fields. The three-waves have negligible mutual dephasing inside the nonlinear ML-TMD (with quadratic surface conductivity $\sigma_{nm} \neq 0$)
as $\ell \simeq 0$; this enables phase-matching free parametric coupling. {\bf c.} Sketch of the geometry of MX$_2$ ML-TMDs. Fast modulation is enabled by extrinsic doping through a gate voltage with gold contacts applied between the ML-TMD and the Bragg mirror.}
\label{fig1}
\end{figure*}
\begin{figure*}
\includegraphics[width=1\textwidth]{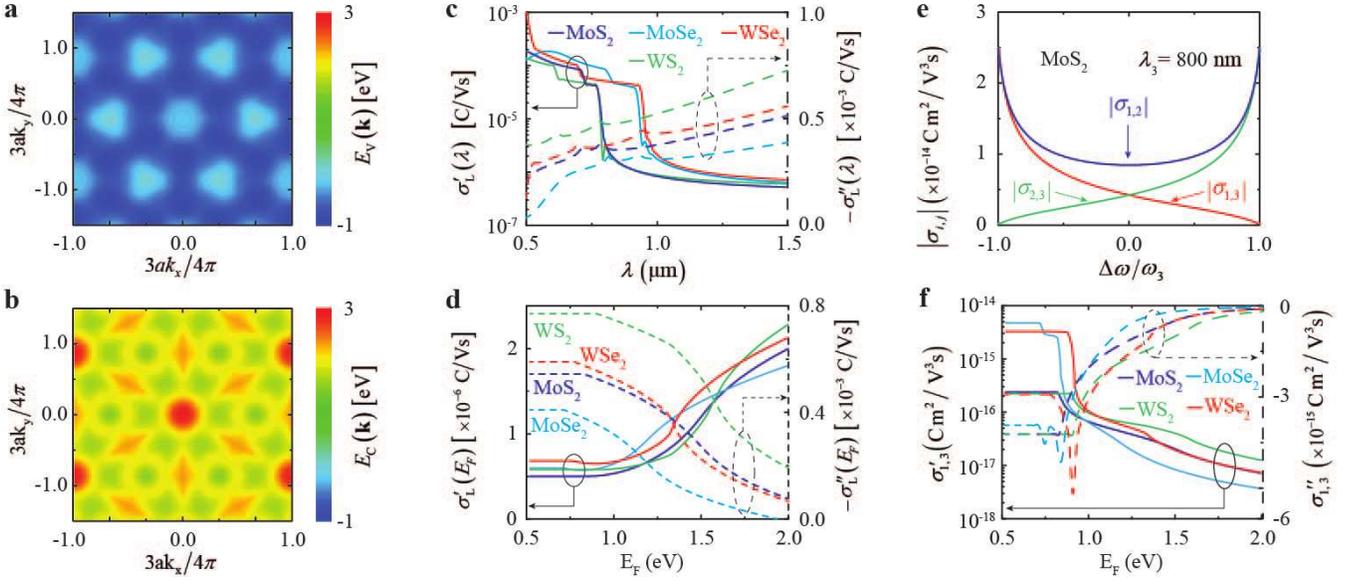}
\caption{{\bf Electronic and optical properties of MX$_2$.} {\bf a,b.} Valence $E_{\rm V}({\bf k})$ and conduction $E_{\rm C}({\bf k})$ energy bands of MoS$_2$, where ${\bf k}$ is the electron wave-vector and $a = 3.19 \r{A}$ is the lattice parameter. {\bf c,d.} Dependence of the linear surface conductivities of ML-TMDs on ({\bf c}) the vacuum wavelength $\lambda$ (for intrinsic doping $E_{\rm F} = 0$) and on ({\bf d}) the Fermi level $E_{\rm F}$ ensuing from extrinsic doping (at $\lambda = 1.6$ $\mu$m). {\bf e}. PDC mixing surface conductivities of MoS$_2$ at $\lambda_3 = 800$ nm as a function of the angular frequency mismatch of down-converted signal and idler waves $\Delta\omega = \omega_1 - \omega_2$ rescaled to the pump angular frequency $\omega_3$. {\bf f}. Dependence of the real and imaginary parts of the PDC mixing conductivity $\sigma_{1,3}$ of MX$_2$ ML-TMDs on the Fermi level $E_{\rm F}$ for $\lambda_1=\lambda_2 = 1.6$ $\mu$m, and $\lambda_3 = 0.8$ $\mu$m.}
\label{fig2}
\end{figure*}
\begin{figure*}
\includegraphics[width=1\textwidth]{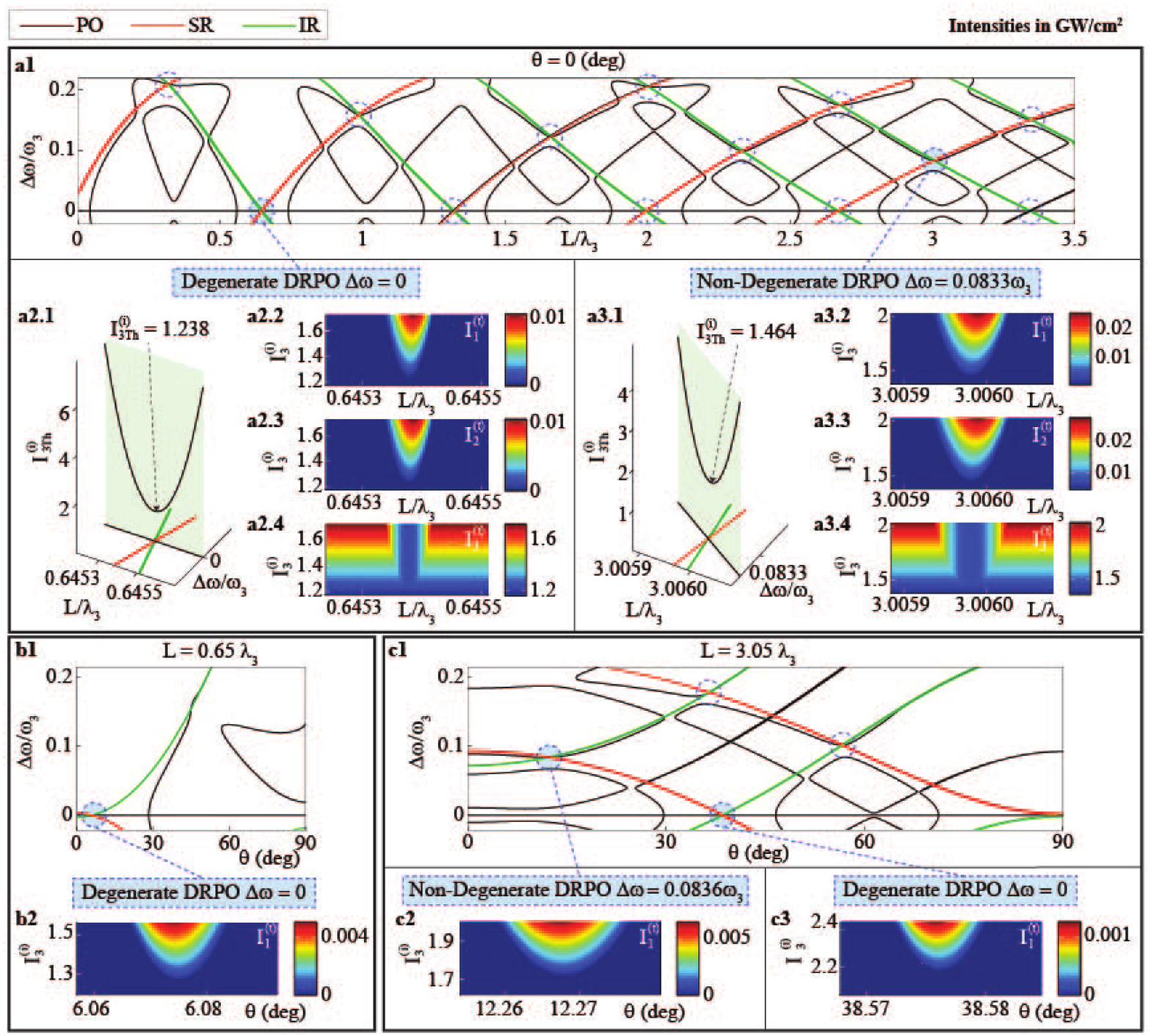}
\caption{{\bf Parametric Oscillations}. Analysis of the doubly resonant parametric oscillations (DRPOs) of a cavity (with PMMA as cavity dielectric) illuminated by a  $\lambda_3 = 780$ nm pump with micron-sized Bragg Mirrors whose stop bad is centered at $1560$ nm. In subfigure {\bf a} the cavity length is used as tuning parameter for normal incidence $\theta = 0$ whereas in {\bf b} and {\bf c} the incidence angle $\theta$ is the tuning parameter for two assigned cavity lengths. {\bf a1.} Identification of DRPOs as intersection among the parametric oscillation (PO) curve and the signal resonance (SR) and idler resonance (IR) curves in the ($L/\lambda_3,\Delta \omega / \omega_3$) plane. {\bf a2} Intensity analysis of the degenerate ($\Delta \omega =0$) DRPO located at $L \simeq 0.645 \lambda_3$ comprising the plots of ({\bf a2.1}) the intensity threshold $I_{3Th}^{(i)}$ versus $L/ \lambda_3$ and $\Delta \omega / \omega_3$ and ({\bf a2.2-4}) the intensities $I_1^{(t)},I_2^{(t)},I_3^{(t)}$ of the transmitted signal, idler and pump fields as functions of the scaled cavity length $L/\lambda_3$ and the incident pump intensity $I_{3}^{(i)}$. {\bf a3} Intensity analysis of the non-degenerate ($\Delta \omega \neq 0$) DRPO located at $L \simeq 3 \lambda_3$ (the panels are the analogous of those of {\bf a2}). {\bf b1.} DRPOs analysis on the ($\theta,\Delta \omega / \omega_3$) plane for $L$ as in {\bf a2}. Note that the degenerate DRPO occurs at a small angle $\theta$. {\bf b2.} Signal intensity analysis of the DRPO showing a feasible $\theta$ range. {\bf c1.} DRPOs analysis on the ($\theta,\Delta \omega / \omega_3$) plane for $L$ as in {\bf a3} revealing a variety of DRPOs ad different angles $\theta$. {\bf c2,c3} signal intensity analysis of two DRPOs identified in {\bf c1}.}
\label{fig3}
\end{figure*}

Since three-wave parametric coupling is intrinsically weak, one can achieve low oscillation thresholds only by doubly or triply resonant optical cavities. In addition, parametric effects are severely hampered by the destructive interference among the three waves propagating with different wavenumbers $k_{1,2,3}$ in the dispersive nonlinear medium because the momentum mismatch $\Delta k = k_3-k_2-k_1$ does not generally vanish (see Fig.1a). To avoid this highly detrimental effect, the use of phase-matching (PM) strategies is imperative. The commonly-adopted birefringence-PM method \cite{Giordmaine1962} is critically sensible to the nonlinear medium orientation. Quasi-PM \cite{Vodopyanov2004,Canalias2007} exploits the momentum due to a manufactured long-scale periodic reversal of the sign of the nonlinear susceptibility and cannot be easily applied in miniaturized system. In semiconductors, PM is achieved by the S-shaped energy-momentum polariton dispersion in the strong coupling of excitons and photons \cite{Savvidis2000,Ciuti2000}, only accessible at low temperatures and large pump angles. Cavity PM \cite{Xie2011}, also denoted ``relaxed'' PM \cite{Clément2015}, occurs in Fabry-Perot microcavities with cavity length $\ell$ shorter than the coherence length $\pi / \Delta k$; this technique drastically reduces the effective quadratic susceptibility $\chi_{\rm eff}^{(2)}$ (see Fig.1a).
Any of the above mentioned PM techniques entails a non-trivial setup design that is further constrained by the need of resonance operation.

In this manuscript, we show that emerging two-dimensional (2D) materials with high quadratic nonlinearity open unprecedented possibilities for tunable parametric micro-sources. Very remarkably, when illuminated with different visible and infrared waves, these novel 2D materials provide a negligible dispersive dephasing owing to their atomic-scale thickness (i.e $\ell \simeq 0$, see Fig.1b). Due to the lack of destructive interference, 2D materials support PDC without any need of satisfying a PM condition. Furthermore, these ``phase-matching-free'' devices turn out to be very versatile and compact, with the additional tunability offered by electrical gating of 2D materials, which provides ultrafast electrical-modulation functionality.

The most famous 2D material, graphene, is not the best candidate for PDC owing to the centrosymmetric structure. In principle, a static external field may break centrosymmetry and induce a $\chi_{\rm eff}^{(2)}$, but
the spectrally-flat absorption of graphene is severely detrimental for PDC. Recent years have witnessed the rise of transition metal dichalcogenides (TMDs) as promising photonic 2D materials. TMDs possess several unusual optical properties dependent on the number of layers. Bulk TMDs are semiconductors with an indirect bandgap, but the optical properties of their monolayer (ML) counterpart are characterized by a direct bandgap ranging from $\sim$ 1.55 eV to $\sim$ 1.9 eV \cite{Mak2010,Splendiani2010,Wang2012} that is beneficial for several optoelectronic applications \cite{Sun2016}. In addition, ML-TMDs have broken centrosymmetry and thus undergo second-order nonlinear processes \cite{Kumar2013,Li2013,Malard2013,Janisch2014,Le2016}. Here we study PDC in micro-cavities embedding ML-TMDs; we find that the cavity design is extremely flexible if compared to standard parametric oscillators thanks to their phase-matching-free operation (see Figs.1a,1b). We demonstrate that, at conventional infrared pump intensity, parametric oscillation occurs in wavelength-sized micro-cavities with ML-TMDs. We show that the output signal and idler frequencies can be engineers thanks to the mode selectivity of doubly-resonant cavities; these frequencies are tuned by the pump incidence angle and modulated electrically by an external gate voltage.

\section{Results}
Two hexagonal lattices of chalcogen atoms embedding a plane of metal atoms arranged at trigonal prismatic sites between the chalcogen neighbors form the structure of ML-TDMs. \cite{Wang2012} Figure~1c shows the lattice structure of MX$_2$ ML-TMDs (M $=$ Mo, W, and X $=$ S, Se), and Figs.2a and 2b report the valence and conduction bands of MoS$_2$ obtained from tight-binding calculations \cite{LWY2013,EPAPS}. The electronic band structure of other MX$_2$ materials considered is qualitatively similar. The direct bandgap is about $1.5$~eV and implies transparency for infrared radiation; the linear surface conductivity has very small real part (corresponding to absorption) and higher imaginary part
at infrared wavelengths. Figure~2c shows the wavelength dependence of the linear surface conductivities of MX$_2$. In the presence of an external pump field with angular frequency $\omega_3$, the ML-TMD second-order nonlinear processes lead to down-converted signal and idler waves with angular frequencies $\omega_1$ and $\omega_2$, such that $\omega_3 =\omega_1+\omega_2$. Figure~2e illustrates the PDC mixing surface conductivities for MoS$_2$. Both linear and nonlinear conductivities are calculated by a perturbative expansion of the tight-binding Hamiltonian of MX$_2$ [see Methods and Supplementary Information (SI)]. For infrared photons with energy smaller than the bandgap, extrinsic doping by an externally applied gate voltage (see Fig.1c) modifies the optical properties and leads to an increase of absorption due to free-carrier collisions and to smaller PDC mixing conductivities. Figures~2d and 2f show the dependence of linear and nonlinear surface conductivities on the Fermi level $E_{\rm F}$. As detailed below, extrinsic doping generally leads to a decrease of PDC efficiency.

Figure~1b shows the parametric oscillator design with ML-TMDs. The cavity consists of a dielectric slab (thickness $L$) surrounded by two Bragg grating mirrors (BGs); the ML-TMD is placed on the left BG inside the cavity. The cavity is illuminated from the left by an incident (i) pump field (frequency $\omega_3$) and the oscillator produces both reflected (r) and transmitted (t) signal and idler fields with frequencies $\omega_1 = (\omega_3 + \Delta \omega)/2$ and $\omega_2 = (\omega_3 - \Delta \omega)/2$, where $\Delta \omega$ is the beat-note frequency of the parametric oscillation (PO).

As detailed in Methods, the cavity equations for the fields do not contain the momentum mismatch $\Delta k$. Indeed, due to their atomic thickness, ML-TMDs are not optically characterized by a refractive index but rather by a surface conductivity. Hence, the parametric coupling produced by the quadratic surface current of ML-TMDs is not hampered by dispersion and no PM condition is required accordingly.
In order to observe signal and idler generation, only the PO condition is required along with the signal resonance (SR) and idler resonance (IR) conditions leading to a dramatic reduction of the intensity threshold (see Methods). Since there is no PM requirement, such requirements can be met by adjusting either the cavity length $L$ or the pump incidence angle $\theta$ as tuning parameters.
For SR and IR, one needs highly reflective mirrors for both signal and idler (see Methods), as obtained by locating stop band of the micron-sized BGs at the half of the pump frequency $\omega_3/2$ \cite{EPAPS}.
Figure~3 shows the PO analysis for a cavity composed of two BGs with polymethyl methacrylate (PMMA) and MoS$_2$ deposited on the left mirror. The infrared pump has wavelength $\lambda_3 = 780$ nm in the spectral region where the nonlinear properties of MoS$_2$ are very pronounced (see Fig.2e). The BGs are tuned with their stop bands centered at $1560$ nm ($=2 \lambda_3$) \cite{EPAPS}. In Fig.3a we consider the case of normal incidence $\theta = 0$ and we plot the PO (black), SR (red), and IR (green) curves in the ($L/\lambda_3,\Delta \omega / \omega_3$)  plane. Doubly resonant POs (DRPOs) corresponding to the intersection points of these three curves \cite{EPAPS} are labeled by dashed circles. Therefore, at pump normal incidence, degenerate ($\Delta \omega =0$) and non-degenerate ($\Delta \omega \neq 0$) DRPOs exist at specific cavity lengths. Note that such oscillations also occur for sub-wavelength cavity lengths ($L < \lambda_3$). Each oscillation starts when the incident pump intensity $I_{3}^{(i)}$ is larger than a threshold $I_{3Th}^{(i)}$ (see Methods) \cite{EPAPS}. Figures~3a2 and 3a3 show the threshold for two specific degenerate and non-denegenerate DRPOs.Panels a2.1 and a3.1 of Fig.3 report the thresholds (black curves on the shadowed vertical planes) corresponding to the PO (black) curves; one can observe that the minimum thresholds occur at SR and IR (identified by the intersection between red and green curves). The minimum intensity thresholds are of the order of GW$/$cm$^2$ and the non-degenerate DRPO threshold is greater than the degenerate DRPO one because the reflectivity of the Bragg mirror is maximum at $\Delta \omega =0$ (i.e. at half the pump frequency, as discussed above). In panels a2.2, a2.3 and a2.4 of Fig. 3 (and, seemingly, panels a3.2, a3.3 and a3.4) we report the basic DRPO features by plotting the intensities $I_1^{(t)},I_2^{(t)},I_3^{(t)}$ of the transmitted signal, idler and pump fields as functions of the scaled cavity length $L/\lambda_3$ and the incident pump intensity.
Note that, in the considered example, the range of $L/\lambda_3$ where the oscillation actually occurs is rather narrow owing to the adopted BGs high reflectivity.

We emphasize that tuning of the PO may be realized by the pump incidence angle $\theta$, which negligibly affects the oscillation thresholds. In Fig.3b and 3c, we analyze the DRPOs by using $\theta$ as tuning parameter for a given cavity length. In particular, in Fig.3b we consider a cavity with fixed length as in Fig.3a2.
The PO, SR and IR curves of Fig.3b1 intersect at a degenerate DRPO point at $\theta \simeq 6$~deg.
In Fig.3b2 we plot the transmitted signal intensity $I_1^{(t)}$ as a function of the pump incidence angle and intensity $I_3^{(i)}$;
one can observe that the intensity threshold is comparable to the case in Fig.3a2 and the range of angles $\theta$ where PO occurs is of the order of a hundredth of degree and experimentally feasible.
We show similar results in Figs.3c1 and 3c2, where the non-degenerate DRPO of Fig.3a3 is investigated in a cavity with slightly different length, and achieved at a finite incident angle with unchanged note-beat frequency $\Delta \omega$. A more accurate analysis of Fig.3c1 also reveals that, for a given $L$, the cavity sustains multiple DRPOs (both degenerate and non-degenerate) at different incidence angles $\theta$.
In Fig.3c3 we plot the transmitted intensity of a degenerate DRPO that grows with the pump intensity above the ignition threshold.

The novel PO with ML-TMDs as nonlinear media are PM-free because of the atomic size of ML-TMDs. The reported several examples of POs with MoS$_2$ can be also designed by other families of ML-TMDs leading to qualitatively similar results. In the Supplementary Material, we compare the calculated dependence of the pump intensity threshold as function of wavelength $\lambda_3$ for parametric oscillators embedding MoS$_2$, WS$_2$ and MoSe$_2$, WSe$_2$; we find that the chosen material affects the minimal threshold intensity in a given spectral range. One can optimize the choice of the material for a desired spectral content and threshold level.

A further degree of freedom offered by ML-TMDs lies in the electrical tunability through an external gate voltage, as depicted in Fig.1c. The gate voltage increases the Fermi level, and hence affects nonlinearity and absorption because of the electron-electron collision in the conduction band (see Figs.2d,f). Although electrical tunability of MX$_2$ has not been hitherto experimentally demonstrated, to the best of our knowledge, we emphasize that such a further degree of freedom is absent in traditional parametric oscillators. In the Supplementary Material, we report the pump intensity threshold as a function of the Fermi level of MoS$_2$, and we show that the threshold may increase by one order of magnitude. The external gate voltage can switch-off PO at fixed optical pump, and fast electrical modulation of the output signal and idler fields can be achieved.

\section{Conclusions}
POs can be excited in micron-sized cavities embedding ML-TMDs as nonlinear media at conventional pump intensities in a PM-free regime.
The cavity design remains inherently free of the complexity imposed by the need for PM and may result into doubly resonant PDC of signal and idler waves.
The flexibility offered by such a novel oscillator design enables the engineering of selective degenerate or non-degenerate down-converted excitations by simply modifying the incident angle of the pump field.
Furthermore, the electrical tunability of ML-TMDs can modulate fast output signal and idler waves by bringing POs below threshold.
Based on our calculations, we envisage that novel parametric oscillators embedding ML-TMDs are a new technology for all the applications in which highly-miniaturized tunable source are relevant, including enviromental detection, security, biophyics, imaging and spectroscopy. PM-free ML-TMD microresonators may also potentially boost the realization of micrometric sources of entangled photons when pumped slightly below threshold, thus paving the way for the development of integrated quantum processors.
\acknowledgments
AM acknowledges useful discussions with F. Javier Garc\'{\i}a de Abajo, Emanuele Distante and Ugo Marzolino.
AC and CR thank the U.S. Army International Technology Center Atlantic for financial support (Grant No. W911NF-14-1-0315).
CC acknowledges funding from the Templeton foundation (Grant number 58277 ) and  PRIN NEMO (reference 2015KEZNYM).

\

{\bf Author Contributions}
A.C. and A.M. conceived the idea and worked out the theory. All the authors discussed the results and wrote the paper.

\section{Methods}

{\bf Parametric down-conversion of MX$_2$}. We calculate the linear and PDC mixing surface conductivities of MX$_2$
starting from the tight-binding (TB) Hamiltonian of the electronic band structure \cite{LWY2013}. Since the properties of infrared photons with energies smaller than the bandgap
are determined by small electron momenta around the K and K' valleys, we approximate the full TB Hamiltonian
as a sum of ${\bf k}\cdot{\bf p}$ Hamiltonians of first and second order $H_0({\bf k},\tau,s)$ \cite{EPAPS},
where ${\bf k}$ is the electron wavenumber and $\tau$ and $s$ are the valley and spin indexes, respectively.
We then derive the light-driven electron dynamics through the minimal coupling prescription leading to the
time-dependent Hamiltonian $H_0\left[{\bf k}+(e/\hbar){\bf A}(t),\tau,s\right]$,
where $-e$ is the electron charge, $\hbar$ is the reduced Planck constant, and ${\bf A}(t)$ is the radiation potential vector, and we obtain Bloch equations
for the interband coherence and the population inversion. Finally, we solve perturbatively the Bloch equations
of ML-TMDs in the weak excitation limit, obtaining the surface current density ${\bf K}(t)$ after integration over the reciprocal
space
\begin{eqnarray}
{\bf K}(t) & = & {\rm Re} \left\{ \sum_{j = 1}^3 \left [ \hat{\sigma}^{\rm L} (\omega_j) {\bf E}_j {\rm e}^{-i\omega_j t} \right] + \hat{\sigma}^{(1,2)} {\bf E}_1 {\bf E}_2 {\rm e}^{-i\omega_3 t} + \right. \nonumber \\
&& \left.+ \hat{\sigma}^{(1,3)} {\bf E}_1^* {\bf E}_3 {\rm e}^{-i\omega_2 t} + \hat{\sigma}^{(2,3)} {\bf E}_2^* {\bf E}_3
{\rm e}^{-i\omega_1 t} \right\},
\end{eqnarray}
\noindent where $\hat{\sigma}^{\rm L}(\omega_j)$ ($j=1,2,3$) and $\hat{\sigma}^{(l,m)}$ ($l,m=1,2,3$) are the linear and
PDC surface conductivity tensors, respectively. Note that our approach is based on the independent-electron approximation
and is fully justified only for infrared photons far from exciton resonances occurring at photon energies higher than $1.5$~eV \cite{Ugeda2014,Selig2016}.

{\bf Parametric oscillations}. The signal, idler and pump fields, labelled with subscripts $1,2,3$ respectively, have frequencies $\omega_n$ satisfying $\omega_1+\omega_2 = \omega_3$. By the Transfer Matrix approach, the full electromagnetic analysis of the cavity (see Supplementary Material) yields the equations
\begin{eqnarray} \label{cavity}
 \Delta _1 Q_1  + \tilde \sigma _{23} Q_2^* Q_3  &=& 0, \nonumber \\
 \Delta _2 Q_2  + \tilde \sigma _{13} Q_1^* Q_3  &=& 0, \nonumber \\
 \Delta _3 Q_3  + \tilde \sigma _{12} Q_1 Q_2  &=& P_3,
\end{eqnarray}

where $Q_1,Q_2,Q_3$ are complex amplitudes proportional to the output fields produced by the pump field which is proportional to the amplitude $P_3$. Here $\tilde \sigma _{nm}$ are scaled quadratic conductivities of the MX$_2$ ML-TMD and
\begin{eqnarray} \label{Delta}
 \Delta _n  &=& \tilde \sigma _n  - \frac{c }{\omega_n} q_n  \left( {\frac{{r_n^{\left( R \right)}  - 1}}{{r_n^{\left( R \right)}  + 1}} + \frac{{r_n^{\left( R \right)} e^{iq_n L}  - e^{ - iq_n L} }}{{r_n^{\left( R \right)} e^{iq_n L}  + e^{ - iq_n L} }}} \right).
\end{eqnarray}
are parameters characterizing the linear cavity where $\tilde \sigma_n$ are scaled linear surface conductivities, $q_n  = \frac{\omega_n}{c} \sqrt {\varepsilon \left( {\omega _n } \right) - \sin ^2 \theta }$ are the longitudinal wavenumbers inside the dielectric slab, $\epsilon(\omega)$ is the relative permittivity of the dielectric slab, $\theta$ is the pump incidence angle whereas $r_n^{\left( R \right)}$ are the complex reflectivities for right illumination of the left Bragg mirror (with vacuum and the dielectric slab on its left and right sides, respectively). It is worth stressing that the phase-mismatch $\Delta k = k_3 - k_1 - k_2$ does not appear in the basic cavity equations (\ref{cavity}). Hence parametric coupling is here not affected by the fields destructive interference and the phase-matching constraint is strictly avoided. Paramateric oscillations (POs) are solutions of Eqs.(\ref{cavity}) with $Q_1 \neq 0$ and $Q_2 \neq 0$ and in this case the compatibility of the first two equations yields (see Supplementary Material)
\begin{equation} \label{threshold}
\left| {P_3 } \right| ^2 \ge \frac{{\Delta _1 \Delta _2^* }}{{\tilde \sigma _{23} \tilde \sigma _{13}^* }}\left| {\Delta _3 } \right|^2,
\end{equation}
which is the leading PO condition. As the right hand side of Eq.~(\ref{threshold}) is generally a complex number, for the PO we have the condition
\begin{equation} \label{necessary}
{\rm arg} \left(\frac{\Delta _1}{\tilde \sigma _{23}}\right) = {\rm arg} \left(\frac{\Delta _2}{\tilde \sigma _{13}}\right).
\end{equation}
Eq.~(\ref{necessary}) can be physically interpreted as a locking of the phase difference ${\arg Q_1} - {\arg Q_2^*}$ allowing the signal and idler to oscillate. Once Eq.(\ref{necessary}) is satisfied, Eq.(\ref{threshold}) provides the pump threshold for the onset of PO. Due to the absolute smallness of the nonlinear surface conductivities, in order to have a feasible threshold, the cavity parameters $|\Delta_n|$ must be minimized. This can be obtained by choosing the doubly resonant condition for signal and idler corresponding to the minima of $|\Delta_1|$ and $|\Delta_2|$, respectively. In order for this minima to be very small, we need that  $|r_1^{\left( R \right)}|$ and $|r_2^{\left( R \right)}|$ are very close to one. One can satisfy such a constraint by a suitable Bragg mirror design to have the stop-band centered at the half of the pump frequency $\omega_3/2$ since, in this case, signal and idler fields experience large mirror reflectance.

\appendix

Here we provide additional information on technical aspects of the theoretical methods used to model parametric down-conversion and the resulting phase-matching free resonant oscillations within the micro-cavities described in the main paper.

\section{Parametric down-conversion of MX$_2$}

We calculate the linear and parametric down-conversion (PDC) mixing surface conductivities of monolayer (ML) transition metal dichalcogenides (TMDs) MX$_2$
(M $=$ Mo, S, and X $=$ S, Se) starting from a tight-binding (TB) description of the electronic band structure of these materials \cite{LWY2013} and
studying the light-driven electron dynamics by means of Bloch equations for the valence and conduction bands. Note that our approach is based on the
independent-electron approximation and is fully justified at the frequencies considered in the main paper since they are far from exciton resonances
happening at photon energies of $\simeq 1.5$ eV or higher \cite{UBS2014,CCB2016}. In addition, for the infrared photon energies
considered, the relevant valence and conduction band regions affecting the infrared response are the ones closer to the band gap around the
K and K' band edges, for which the full TB Hamiltonian can be approximated by a two-band ${\bf k}\cdot{\bf p}$ Hamiltonian
${\cal H}_0({\bf k},\tau,s) = {\cal H}_1({\bf k},\tau,s) + {\cal H}_2({\bf k},\tau)$ \cite{LWY2013}, where
\begin{eqnarray}
&&{\cal H}_1({\bf k},\tau,s) = \left[ \begin{array}{cc} \Delta/2 & t_0 a (\tau k_x - i k_y) \\
t_0 a (\tau k_x + i k_y) & \tau s \Lambda - \Delta/2 \end{array} \right], \\
&& {\cal H}_2({\bf k},\tau,s) = \left[ \begin{array}{cc} \gamma_1a^2k^2 & \gamma_3a^2(\tau k_x + i k_y)^2 \\
\gamma_3a^2(\tau k_x - i k_y)^2 &  \gamma_2a^2k^2 \end{array} \right], \\
\end{eqnarray}
and $\tau,s=\pm 1$ label non-degenerate valleys and spins, while ${\bf k} = (k_x~k_y)$ indicates the electron wave-vector.
\begin{table}[b]
\renewcommand{\thetable}{S\arabic{table}}
	\caption{Fitted constants of the the two-band ${\bf k}\cdot{\bf p}$ Hamiltonian ${\cal H}_0({\bf k},\tau,s)$.}
	\begin{tabular}{llllll}
		\hline
		Constants       & ~~MoS$_2$  & ~~MoSe$_2$  & ~~WS$_2$  & ~~WSe$_2$     \\
		\hline
		~~$a$ (\r{A})     & ~~$3.190$  & ~~$3.326$   & ~~$3.191$ & ~~$3.325$     \\
		~~$\Delta$   (eV) & ~~$1.658$  & ~~$1.429$   & ~~$1.806$ & ~~$1.541$     \\
		~~$t_0$      (eV) & ~~$0.933$  & ~~$0.768$   & ~~$1.196$ & ~~$1.016$     \\
		~~$\Lambda$  (eV) & ~~$0.073$  & ~~$0.091$   & ~~$0.211$ & ~~$0.228$     \\
		~~$\gamma_1$ (eV) & ~~$0.351$  & ~~$0.291$   & ~~$0.443$ & ~~$0.404$     \\
		~~$\gamma_2$ (eV) & $-0.198$  & $-0.191$   & $-0.211$ & $-0.216$     \\
		~~$\gamma_3$ (eV) & $-0.143$  & $-0.099$   & $-0.199$ & $-0.150$     \\
		\hline
	\end{tabular}
\end{table}
The physical parameters of ${\cal H}_0({\bf k},\tau,s)$ are obtained by fitting the ${\bf k}\cdot{\bf p}$ valence and conduction energy bands with the ones
obtained from first-principles GW simulations \cite{LWY2013} accounting for both non-degenerate valleys and spin-orbit coupling, and are listed in Table\ S1.
We calculate the linear and PDC mixing conductivities of ML-TMDs by introducing the time-dependent Hamiltonian ${\cal H}_0(\mbox{\boldmath${\kappa}$}(t),\tau,s)$,
where we have replaced the electron wave-vector with the minimum coupling prescription for the electron quasi-momentum
$\hbar\mbox{\boldmath${\kappa}$}(t) = \hbar{\bf k} + e {\bf A}(t)$, where $e$ is the electron charge and ${\bf A}(t)$ is the electromagnetic potential vector
accounting for pump, signal, and idler waves. With this prescription, we define unperturbed and interacting Hamiltonians ${\cal H}_0({\bf k},\tau,s)$ and ${\cal H}_{\rm I}({\bf k},\tau,s,t)$, respectively, and write the total Hamiltonian as ${\cal H}_{\rm T}({\bf k},\tau,s,t) = {\cal H}_0[\mbox{\boldmath${\kappa}$}(t),\tau,s] = {\cal H}_0({\bf k},\tau,s) + {\cal H}_{\rm I}({\bf k},\tau,s,t)$, where
\begin{eqnarray}
{\cal H}_{\rm I}({\bf k},\tau,s,t) & = & \frac{e}{\hbar} [D_x A_x(t) + D_y A_y(t)] + \frac{e^2}{\hbar^2} [D_{xx} A_x^2(t) + D_{xy} A_x(t)A_y(t) + D_{yy} A_y^2(t)],
\end{eqnarray}
and the interaction operators are explicitly given by
\begin{eqnarray}
&& D_x     = t_0a\tau \left[ |\psi_{\rm V}\rangle\langle\psi_{\rm C}| + |\psi_{\rm C}\rangle\langle\psi_{\rm V}| \right], \nonumber \\
&& D_y     = i t_0 a  \left[ |\psi_{\rm V}\rangle\langle\psi_{\rm C}| - |\psi_{\rm C}\rangle\langle\psi_{\rm V}| \right], \nonumber \\
&& D_{xx}  = \gamma_1a^2 |\psi_{\rm C}\rangle\langle\psi_{\rm C}| + \gamma_2a^2 |\psi_{\rm V}\rangle\langle\psi_{\rm V}|
+ \gamma_3a^2 \left[ |\psi_{\rm V}\rangle\langle\psi_{\rm C}| + |\psi_{\rm C}\rangle\langle\psi_{\rm V}| \right], \nonumber \\
&& D_{yy}  = \gamma_1a^2 |\psi_{\rm C}\rangle\langle\psi_{\rm C}| + \gamma_2a^2 |\psi_{\rm V}\rangle\langle\psi_{\rm V}|
- \gamma_3a^2 \left[ |\psi_{\rm V}\rangle\langle\psi_{\rm C}| + |\psi_{\rm C}\rangle\langle\psi_{\rm V}| \right], \nonumber \\
&& D_{xy}  = 2i\gamma_3a^2\tau \left[ |\psi_{\rm C}\rangle\langle\psi_{\rm V}| - |\psi_{\rm V}\rangle\langle\psi_{\rm C}| \right]. \nonumber \\
\end{eqnarray}
In the expressions above we use the Dirac notation for the conduction $|\psi_{\rm C}\rangle$ and valence $|\psi_{\rm V}\rangle$ band eigenstates,
and we approximate the matrix elements by their values at the band edges (${\bf k} = 0$). Inserting the Ansatz
$|\psi\rangle = c_-|\psi_{\rm V}\rangle + c_+ |\psi_{\rm C}\rangle$ in the time-dependent Schr\"odinger equation
$i\hbar\partial_t |\psi\rangle = {\cal H}_{\rm T} |\psi\rangle$, and defining the inversion population $n_{\bf k}=|c_+|^2-|c_-|^2$ and the interband
coherence $\rho_{\bf k}=c_+c_-^*$, one gets
\begin{eqnarray}
\dot{\rho}_{\bf k} & = & -\frac{i}{\hbar} (E_{\rm C} - E_{\rm V})\rho_{\bf k} - \gamma\rho_{\bf k} + \frac{ie}{\hbar^2} n_{\bf k}\left\{ D_x^{\rm CV} A_x(t) + D_y^{\rm CV} A_y(t) + \frac{e}{\hbar} [ D_{xx}^{\rm CV} A_x^2(t) + D_{xy}^{\rm CV} A_x(t)A_y(t) + D_{yy}^{\rm CV} A_y^2(t) ] \right\} + \nonumber \\
&&  +  \frac{ie^2}{\hbar^3} \left[ (D_{xx}^{\rm VV} - D_{xx}^{\rm CC})A_x^2(t) + (D_{yy}^{\rm VV} - D_{yy}^{\rm CC})A_y^2(t) \right] \rho_{\bf k}, \nonumber \\
\dot{n}_{\bf k} & = & - \frac{4e}{\hbar^2} {\rm Im} \left\{ \rho_{\bf k} \left[ D_{x}^{\rm VC}A_x(t) + D_{y}^{\rm VC}A_y(t) + \frac{e}{\hbar} [ D_{xx}^{\rm VC}A_x^2(t) + D_{xy}^{\rm VC}A_x(t)A_y(t) + D_{yy}^{\rm VC}A_y^2(t)] \right]\right\},
\end{eqnarray}
where $E_{\rm C} ({\bf k})$ and $E_{\rm V} ({\bf k})$ are the conduction and valence energy bands of the unperturbed Hamiltonian ${\cal H}_0$,
$D^{\rm CV}_j=\langle \psi_{\rm C} | D_j | \psi_{\rm V}\rangle$ are the interaction matrix elements, and we have introduced a phenomenological relaxation
rate $\gamma = 10$ ps$^{-1}$ accounting for coherence dephasing \cite{YYC2014}. In order to obtain the PDC surface conductivities, we consider a coherent superposition
of three monochromatic fields
${\bf E}(t) = {\rm Re} \left\{{\bf E}_1{\rm e}^{-i\omega_1 t}+{\bf E}_2{\rm e}^{-i\omega_2 t}+{\bf E}_3{\rm e}^{-i\omega_3 t}\right\}$
with amplitudes ${\bf E}_1$, ${\bf E}_2$, and ${\bf E}_3$, and with angular frequencies $\omega_1$, $\omega_2$, and $\omega_3$, respectively.
In our notation, the field ${\bf E}_3$ indicates the external pump field, while ${\bf E}_1$, ${\bf E}_2$ label the down-converted signal and idler
fields, respectively. The down-converted angular frequencies $\omega_1$ and $\omega_2$ are not independent, but are such that $\omega_1+\omega_2=\omega_3$
owing to energy conservation. The electromagnetic potential vector related to the coherent superposition of pump, signal, and idler waves is thus given by
\begin{equation}
{\bf A}(t) = {\rm Re} \left\{({\bf E}_1/i\omega_1){\rm e}^{-i\omega_1 t}+({\bf E}_2/i\omega_2){\rm e}^{-i\omega_2 t}+({\bf E}_3/i\omega_3){\rm e}^{-i\omega_3 t}\right\}.
\end{equation}
We then solve perturbatively the equations above in the vanishing temperature $T\rightarrow 0$ and weak excitation limits such that
$n_{\bf k}\approx - \Theta \left[ E_{\rm C}({\bf k}) - E_{\rm F} \right]$, where $\Theta(x)$ indicates the Heaviside step function and $E_{\rm F}$ is the Fermi energy. Taking the Ansatz
$\rho_{\bf k} = \sum_{j = \pm1,\pm2,\pm3} \rho_{|j|}^{(j/|j|)}{\rm e}^{i(j/|j|)\omega_{|j|} t}$ and disregarding generation of higher harmonics we obtain analytical expressions for the coefficients $\rho_{|j|}^{(j/|j|)}$, finding that the macroscopic surface current density given by
\begin{eqnarray}
{\bf K}(t) & = & - \frac{e}{4\pi^2\hbar}\sum_{\tau,s=-1,1} \int_{-\infty}^{+\infty} dk_x \int_{-\infty}^{+\infty} dk_y \left[ \langle \psi(t)| \nabla_{\bf k} {\cal H}_{\rm T} (t) | \psi (t) \rangle - \langle \psi_{\rm V}| \nabla_{\bf k} {\cal H}_{\rm T} (t) | \psi_{\rm V}  \rangle \right] = \\
           & = & - \frac{e}{2\pi^2\hbar}\sum_{\tau,s=-1,1} \int_{-\infty}^{+\infty} dk_x \int_{-\infty}^{+\infty} dk_y {\rm Re} \left\{ \rho_{\bf k} (t)\left[ \nabla_{\bf k} {\cal H}_0^{\rm VC} + \nabla_{\bf k} D_x^{\rm VC} \frac{e}{\hbar} A_x(t) + \nabla_{\bf k} D_y^{\rm VC} \frac{e}{\hbar} A_y(t) + \right. \right. \nonumber \\
					 &   & \left. \left. + \nabla_{\bf k} D_{xx}^{\rm VC} \frac{e^2}{\hbar^2} A_x^2(t) + \nabla_{\bf k} D_{xy}^{\rm VC} \frac{e^2}{\hbar^2} A_x(t)A_y(t) + \nabla_{\bf k} D_{yy}^{\rm VC} \frac{e^2}{\hbar^2} A_y^2(t) \right] + \right. \nonumber \\
					 &   & \left. + \Theta \left[ E_{\rm F} - E_{\rm C}({\bf k}) \right] \left[D_{xx}^{\rm CC}\frac{e}{\hbar}A_x(t)+D_{yy}^{\rm CC}\frac{e}{\hbar}A_y(t)\right]\right\}, \nonumber
\end{eqnarray}
can be recast into
\begin{equation}
{\bf K}(t) = {\rm Re} \left\{ \sum_{j = 1}^3 \left [ \hat{\sigma}^{\rm L} (\omega_j) {\bf E}_j {\rm e}^{-i\omega_j t} \right] + \hat{\sigma}^{(1,2)} {\bf E}_1
{\bf E}_2 {\rm e}^{-i\omega_3 t} + \hat{\sigma}^{(1,3)} {\bf E}_1^* {\bf E}_3 {\rm e}^{-i\omega_2 t} + \hat{\sigma}^{(2,3)} {\bf E}_2^* {\bf E}_3
{\rm e}^{-i\omega_1 t} \right\},
\end{equation}
where $\hat{\sigma}^{\rm L}(\omega_j)$ ($j=1,2,3$) and $\hat{\sigma}^{(l,m)}$ ($l,m=1,2,3$) are the linear and PDC surface conductivity tensors, respectively, and we have neglected again generation of higher harmonics. Since centrosymmetry is broken along the $y$-direction
in the notation used, the relevant components of the surface conductivity tensors for PDC are the ones such that pump, signal, and idler fields are polarized
along the $y$-direction, for which
\begin{eqnarray}
&& \sigma^{\rm L}_{yy} (\omega) = \frac{ie^2D_{yy}^{\rm CC}}{2\pi^2\hbar^2(\omega+i\gamma)} \sum_{\tau,s = \pm 1} \int_{-\infty}^{+\infty} d k_x \int_{-\infty}^{+\infty} d k_y \Theta \left[ E_{\rm F} - E_{\rm C}({\bf k}) \right] + \\
&& + \frac{e^2}{4i\pi^2\hbar^2\omega} \sum_{\tau,s = \pm 1} \int_{-\infty}^{+\infty} d k_x \int_{-\infty}^{+\infty} d k_y \Theta \left[ E_{\rm C}({\bf k}) - E_{\rm F} \right] \left\{  \frac{|D_y^{\rm CV}|^2}{[(E_{\rm C} - E_{\rm V}) - \hbar (\omega + i\gamma)]} + \frac{|D_y^{\rm CV}|^2}{[(E_{\rm C} - E_{\rm V}) + \hbar (\omega + i\gamma)]} \right\}, \nonumber \\
&& \sigma^{\rm (1,2)}_{yyy} (\omega_1,\omega_2,\omega_3) = \frac{-e^3}{4\pi^2\hbar^3\omega_1\omega_2} \sum_{\tau,s = \pm 1}\sum_{j = 1}^3 \int_{-\infty}^{+\infty} d k_x \int_{-\infty}^{+\infty} d k_y \Theta \left[ E_{\rm C}({\bf k}) - E_{\rm F} \right] \left\{  \frac{D_y^{\rm CV}D_{yy}^{\rm VC}}{[(E_{\rm C} - E_{\rm V}) - \hbar (\omega_j + i\gamma)]} + \right. \nonumber \\
&& \left. + \frac{D_y^{\rm VC}D_{yy}^{\rm CV}}{[(E_{\rm C} - E_{\rm V}) + \hbar (\omega_j + i\gamma)]}  \right\}, \nonumber \\
&& \sigma^{\rm (1,3)}_{yyy} (\omega_1,\omega_2,\omega_3) = \frac{e^3}{4\pi^2\hbar^3\omega_1\omega_3} \sum_{\tau,s = \pm 1} \int_{-\infty}^{+\infty} d k_x \int_{-\infty}^{+\infty} d k_y \Theta \left[ E_{\rm C}({\bf k}) - E_{\rm F} \right] \left\{ \frac{D_y^{\rm CV}D_{yy}^{\rm VC}}{[(E_{\rm C} - E_{\rm V}) + \hbar (\omega_1 - i\gamma)]} + \right. \\
&& \left. + \frac{D_y^{\rm VC}D_{yy}^{\rm CV}}{[(E_{\rm C} - E_{\rm V}) - \hbar (\omega_1 - i\gamma)]} + \frac{D_y^{\rm CV}D_{yy}^{\rm VC}}{[(E_{\rm C} - E_{\rm V}) + \hbar (\omega_2 + i\gamma)]} + \frac{D_y^{\rm VC}D_{yy}^{\rm CV}}{[(E_{\rm C} - E_{\rm V}) - \hbar (\omega_2 + i\gamma)]} + \right. \nonumber \\
&& \left. + \frac{D_y^{\rm CV}D_{yy}^{\rm VC}}{[(E_{\rm C} - E_{\rm V}) - \hbar (\omega_3 + i\gamma)]} + \frac{D_y^{\rm VC}D_{yy}^{\rm CV}}{[(E_{\rm C} - E_{\rm V}) + \hbar (\omega_3 + i\gamma)]} \right\}, \nonumber \\
&& \sigma^{\rm (2,3)}_{yyy} (\omega_1,\omega_2,\omega_3) = \frac{e^3}{4\pi^2\hbar^3\omega_2\omega_3} \sum_{\tau,s = \pm 1} \int_{-\infty}^{+\infty} d k_x \int_{-\infty}^{+\infty} d k_y \Theta \left[ E_{\rm C}({\bf k}) - E_{\rm F} \right] \left\{ \frac{D_y^{\rm CV}D_{yy}^{\rm VC}}{[(E_{\rm C} - E_{\rm V}) + \hbar (\omega_1 + i\gamma)]} + \right. \\
&& \left. + \frac{D_y^{\rm VC}D_{yy}^{\rm CV}}{[(E_{\rm C} - E_{\rm V}) - \hbar (\omega_1 + i\gamma)]} + \frac{D_y^{\rm CV}D_{yy}^{\rm VC}}{[(E_{\rm C} - E_{\rm V}) + \hbar (\omega_2 - i\gamma)]} + \frac{D_y^{\rm VC}D_{yy}^{\rm CV}}{[(E_{\rm C} - E_{\rm V}) - \hbar (\omega_2 - i\gamma)]} + \frac{D_y^{\rm CV}D_{yy}^{\rm VC}}{[(E_{\rm C} - E_{\rm V}) - \hbar (\omega_3 + i\gamma)]} + \right. \nonumber \\
&& \left. + \frac{D_y^{\rm VC}D_{yy}^{\rm CV}}{[(E_{\rm C} - E_{\rm V}) + \hbar (\omega_3 + i\gamma)]} \right\}. \nonumber
\end{eqnarray}
Data reported in the main paper are obtained through the expressions above. In what follows, for convenience we will assume the simplified notation $\sigma_n = \sigma^{\rm L}_{yy}(\omega_n)$ and $\sigma_{nm} = \sigma^{(n,m)}_{yyy}(\omega_1,\omega_2,\omega_3)$ since the pump, signal, and idler electric fields are polarized in the $y$-direction for maximizing PDC within the micro-cavity.

\section{Equations for the output fields}
\begin{figure}[b]
\center
\includegraphics*[width=0.8\textwidth]{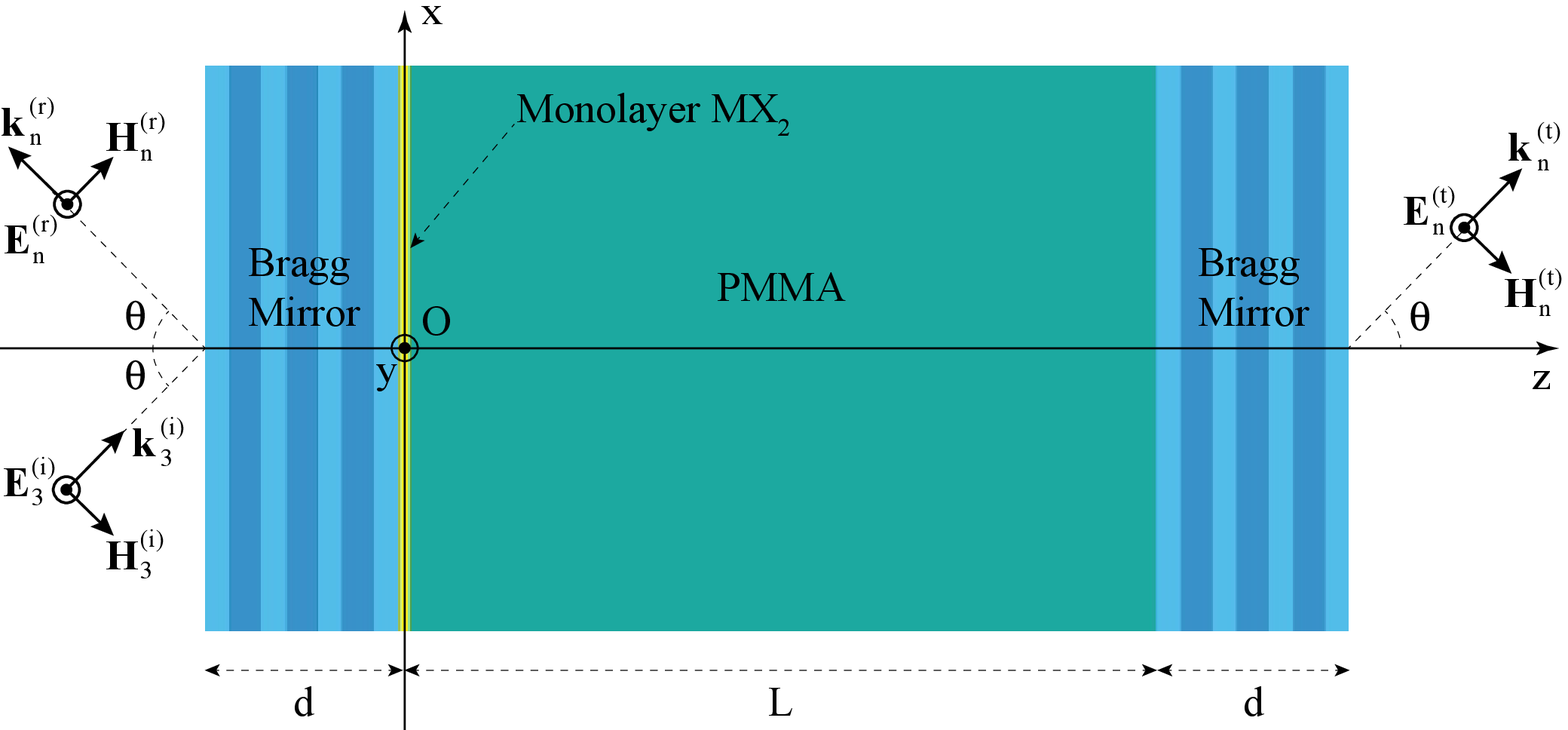}
\caption{Parametric oscillator geometry.}
\label{S1}
\end{figure}

In Fig.\ref{S1} we sketch the geometry of the parametric oscillator (PO) considered in our calculations. A dielectric (PMMA) slab of thickness $L$ with a MX$_2$ monolayer lying on its left side (at $z=0$) is placed between two Bragg mirrors of thickness $d$ (for convenience we choose the right mirror to be the reflected $z \rightarrow -z$ copy of the left one). The left side of the cavity is illuminated with an incident $(i)$ pump field which is a monochromatic Transverse Electric (TE) plane wave of frequency $\omega_3$ with incidence angle $\theta$. In addition to the reflected $(r)$ and transmitted $(t)$ pump fields, due to PDC, the cavity also produces $(r)$ and $(t)$ TE plane waves at the frequencies $\omega_1$ (signal) and $\omega_2$ (idler) such that $\omega_3 = \omega_1 + \omega_2$. It is convenient to set
\begin{eqnarray} \label{frequencies}
\omega_1 &=& \frac{1}{2} \left( \omega_3 +  \Delta \omega \right), \nonumber \\
\omega_2 &=& \frac{1}{2} \left( \omega_3 -  \Delta \omega \right),
\end{eqnarray}

\noindent since the note-beat frequency $\Delta \omega = \omega_1 - \omega_2$ is sufficient to label the signal and idler frequencies produced by a pump field of frequency $\omega_3$. Conservation of transverse momentum of the three fields implies that the their complex amplitudes ($\sim e^{-i\omega_n t}$, $n=1,2,3$) are
\begin{eqnarray} \label{TEfield}
 {\bf{E}}_n  &=& e^{i\frac{\omega_n}{c} x\sin \theta } \left[ {A_{ny} \left( z \right){\bf{\hat e}}_y } \right], \nonumber \\
 {\bf{H}}_n  &=& e^{i\frac{\omega_n}{c} x\sin \theta } \sqrt {\frac{{\varepsilon _0 }}{{\mu _0 }}} \left[ {A_{nx} \left( z \right){\bf{\hat e}}_x  + A_{ny} \left( z \right) \sin \theta {\bf{\hat e}}_z } \right].
\end{eqnarray}
Accordingly the three two-component column vectors $\left(A_{nx}(z) \: A_{ny}(z)\right)^T$ fully describe the field and in vacuum, i.e. outside the cavity, they are
\begin{equation} \label{vacuumfields}
\begin{cases}
\begin{pmatrix}
   {A_{nx} }  \\
   {A_{ny} }  \\
\end{pmatrix} = E_n^{\left( i \right)} \begin{pmatrix}
   { - \cos \theta }  \\
   1  \\
\end{pmatrix}e^{i\frac{\omega_n}{c} \left( {z + d} \right)\cos \theta}  + E_n^{\left( r \right)} \begin{pmatrix}
   {\cos \theta }  \\
   1  \\
\end{pmatrix}e^{ - i\frac{\omega_n}{c} \left( {z + d} \right)\cos \theta }, & z <  - d,  \\
\begin{pmatrix}
   {A_{nx} }  \\
   {A_{ny} }  \\
\end{pmatrix} = E_n^{\left( t \right)} \begin{pmatrix}
   { - \cos \theta }  \\
   1  \\
\end{pmatrix}e^{i\frac{\omega_n}{c} \left( {z - L - d} \right)\cos \theta },  & z> L+d,
\end{cases}
\end{equation}
where $E_n^{\left( i \right)}$, $E_n^{\left( r \right)}$ and $E_n^{\left( t \right)}$ are field amplitudes with $E_1^{\left( i \right)} = E_2^{\left( i \right)} =0$. Resorting to the transfer matrix approach, the fields at the left ($z=0^-$) and right ($z=0^+$) sides of the MX$_2$ monolayer are
\begin{eqnarray} \label{propagation}
 \begin{pmatrix}
   {A_{nx} }  \\
   {A_{ny} }  \\
\end{pmatrix}_{z = 0^ -  }  &=& F_n \begin{pmatrix}
   {A_{nx} }  \\
   {A_{ny} }  \\
\end{pmatrix}_{z =  - d} , \nonumber \\
 \begin{pmatrix}
   {A_{nx} }  \\
   {A_{ny} }  \\
\end{pmatrix}_{z = 0^ +  }  &=& B_n B'_n \begin{pmatrix}
   {A_{nx} }  \\
   {A_{ny} }  \\
\end{pmatrix}_{z = L + d},
\end{eqnarray}
where $F_n$, $B_n$ and $B'_n$ are the transfer matrix describing the forward, backward and backward propagations through the left Bragg mirror, the dielectric slab and the right Bragg mirror, respectively. The transfer matrix $F_n$ of the left Bragg mirror is the (ordered) product of the transfer matrices of the slabs composing the mirror. For later convenience it is useful to represent this matrix as \cite{Kavok}
\begin{eqnarray} \label{matrices1}
F_n  &=& \begin{pmatrix}
   {\frac{{c q_n }}{{\omega_n \cos \theta }}\left[ {\frac{{t_n^{\left( L \right)} t_n^{\left( R \right)}  + \left( {1 + r_n^{\left( L \right)} } \right)\left( {1 - r_n^{\left( R \right)} } \right)}}{{2t_n^{\left( R \right)} }}} \right]} &
{\frac{{c q_n }}{{\omega_n }}\left[ {\frac{{ - t_n^{\left( L \right)} t_n^{\left( R \right)}  + \left( {1 - r_n^{\left( L \right)} } \right)\left( {1 - r_n^{\left( R \right)} } \right)}}{{2t_n^{\left( R \right)} }}} \right]}  \\
   {\frac{1}{{\cos \theta }}\left[ {\frac{{ - t_n^{\left( R \right)} t_n^{\left( L \right)}  + \left( {1 + r_n^{\left( L \right)} } \right)\left( {1 + r_n^{\left( R \right)} } \right)}}{{2t_n^{\left( R \right)} }}} \right]} &
{\left[ {\frac{{t_n^{\left( L \right)} t_n^{\left( R \right)}  + \left( {1 - r_n^{\left( R \right)} } \right)\left( {1 + r_n^{\left( R \right)} } \right)}}{{2t_n^{\left( R \right)} }}} \right]}  \\
\end{pmatrix}, \nonumber \\
\end{eqnarray}
where $q_n  = \frac{\omega_n}{c} \sqrt {\varepsilon \left( {\omega _n } \right) - \sin ^2 \theta }$ are the longitudinal wavenumbers inside the dielectric slab, $\epsilon(\omega)$ is the relative permittivity of the dielectric slab whereas $r_n^{\left( L \right)},t_n^{\left( L \right)},r_n^{\left( R \right)},t_n^{\left( R \right)}$ are the complex reflectivities $r$ and transmittivities $t$ for left $(L)$ and right $(R)$ illumination of the left Bragg mirror (with vacuum and the dielectric on its left and right sides, respectively). The other relevant transfer matrices are \cite{Kavok}
\begin{eqnarray} \label{matrices2}
B_n  &=& \begin{pmatrix}
   {\cos \left( {q_n L} \right)} & {i\frac{{c q_n }}{{\omega_n }}\sin \left( {q_n L} \right)}  \\
   {i{{\frac{\omega_n}{c q_n} }}\sin \left( {q_n L} \right)} & {\cos \left( {q_n L} \right)}  \\
\end{pmatrix}, \nonumber \\
B'_n &=& \begin{pmatrix}
   1 & 0  \\
   0 & { - 1}  \\
\end{pmatrix} F_n \begin{pmatrix}
   1 & 0  \\
   0 & { - 1}  \\
\end{pmatrix},
\end{eqnarray}
where the last of Eqs. (\ref{matrices2}) is a consequence of the fact that the right Bragg mirror is the reflected image of the left one. Using Eqs. (\ref{vacuumfields}), Eqs. (\ref{propagation}) yield
\begin{eqnarray} \label{boundaryfields}
\begin{pmatrix}
   {A_{nx} }  \\
   {A_{ny} }  \\
\end{pmatrix}_{z = 0^ -  } &=&\begin{pmatrix}
   {V_{nx}^{\left( i \right)} }  \\
   {V_{ny}^{\left( i \right)} }  \\
\end{pmatrix}E_n^{\left( i \right)}  +\begin{pmatrix}
   {V_{nx}^{\left( r \right)} }  \\
   {V_{ny}^{\left( r \right)} }  \\
\end{pmatrix}E_n^{\left( r \right)} , \nonumber \\
\begin{pmatrix}
   {A_{nx} }  \\
   {A_{ny} }  \\
\end{pmatrix}_{z = 0^ +  }  &=&\begin{pmatrix}
   {V_{nx}^{\left( t \right)} }  \\
   {V_{ny}^{\left( t \right)} }  \\
\end{pmatrix}E_n^{\left( t \right)},
\end{eqnarray}
where
\begin{eqnarray} \label{amplitudesV}
\begin{pmatrix}
   {V_{nx}^{\left( i \right)} }  \\
   {V_{ny}^{\left( i \right)} }  \\
\end{pmatrix} &=& \frac{1}{{t_n^{\left( R \right)} }}\begin{pmatrix}
   {\frac{{c q_n }}{\omega_n }\left[ { - t_n^{\left( L \right)} t_n^{\left( R \right)}  + r_n^{\left( L \right)} \left( {r_n^{\left( R \right)}  - 1} \right)} \right]}  \\
   {t_n^{\left( L \right)} t_n^{\left( R \right)}  - r_n^{\left( L \right)} \left( {r_n^{\left( R \right)}  + 1} \right)}  \\
\end{pmatrix}, \nonumber \\
\begin{pmatrix}
   {V_{nx}^{\left( r \right)} }  \\
   {V_{ny}^{\left( r \right)} }  \\
\end{pmatrix} &=& \frac{1}{{t_n^{\left( R \right)} }}\begin{pmatrix}
   { - \frac{{c q_n }}{{\omega_n }}\left( {r_n^{\left( R \right)}  - 1} \right)}  \\
   {r_n^{\left( R \right)}  + 1}  \\
\end{pmatrix}, \nonumber \\
\begin{pmatrix}
   {V_{nx}^{\left( t \right)} }  \\
   {V_{ny}^{\left( t \right)} }  \\
\end{pmatrix} &=& \frac{1}{{t_n^{\left( R \right)} }}\begin{pmatrix}
   {\frac{{c q_n }}{\omega_n }\left( {r_n^{\left( R \right)} e^{iq_n L}  - e^{ - iq_n L} } \right)}  \\
   {\left( {r_n^{\left( R \right)} e^{iq_n L}  + e^{ - iq_n L} } \right)}
\end{pmatrix}.
\end{eqnarray}
The monolayer of MX$_2$ in the presence of the above TE electromagnetic field hosts a surface current whose harmonic complex amplitudes are ${\bf K}_n = K_n \hat{\bf e}_y$  where
\begin{eqnarray}
 K_{1}  &=& \left[ \sigma _1 E_{1y}  + \sigma _{23} E_{2y}^* E_{3y} \right]_{z=0}  , \nonumber \\
 K_{2}  &=& \left[ \sigma _2 E_{2y}  + \sigma _{13} E_{1y}^* E_{3y}  \right]_{z=0}, \nonumber \\
 K_{3}  &=& \left[ \sigma _3 E_{3y}  + \sigma _{12} E_{1y} E_{2y}    \right]_{z=0},
\end{eqnarray}
showing both a linear and a quadratic response to the electric field. The effect of such surface current on the field is provided by the electromagnetic boundary conditions at $z=0$, namely ${\bf{\hat e}}_z  \times \left\{ {\left[ {{\bf{E}}_n } \right]_{z = 0^ +  }  - \left[ {{\bf{E}}_n } \right]_{z = 0^ -  } } \right\} = {\bf{0}}$ and ${\bf{\hat e}}_z  \times \left\{ {\left[ {{\bf{H}}_n } \right]_{z = 0^ +  }  - \left[ {{\bf{H}}_n } \right]_{z = 0^ -  } } \right\} = {\bf{K}}_n$ which, using Eqs. (\ref{TEfield}), can be casted within the two-component column vector description as
\begin{eqnarray} \label{matching}
 \begin{pmatrix}
   {A_{1x} }  \\
   {A_{1y} }  \\
\end{pmatrix}_{z = 0^ +  }  - \begin{pmatrix}
   {A_{1x} }  \\
   {A_{1y} }  \\
\end{pmatrix}_{z = 0^ -  }  = \begin{pmatrix}
   {\tilde \sigma _1 A_{1y}  + \tilde \sigma _{23} A_{2y}^* A_{3y} }  \\
   0  \\
\end{pmatrix}_{z = 0^ +  }, \nonumber \\
 \begin{pmatrix}
   {A_{2x} }  \\
   {A_{2y} }  \\
\end{pmatrix}_{z = 0^ +  }  - \begin{pmatrix}
   {A_{2x} }  \\
   {A_{2y} }  \\
\end{pmatrix}_{z = 0^ -  }  = \begin{pmatrix}
   {\tilde \sigma _2 A_{2y}  + \tilde \sigma _{13} A_{1y}^* A_{3y} }  \\
   0  \\
\end{pmatrix}_{z = 0^ +  }, \nonumber \\
 \begin{pmatrix}
   {A_{3x} }  \\
   {A_{3y} }  \\
\end{pmatrix}_{z = 0^ +  }  - \begin{pmatrix}
   {A_{3x} }  \\
   {A_{3y} }  \\
\end{pmatrix}_{z = 0^ -  }  = \begin{pmatrix}
   {\tilde \sigma _3 A_{3y}  + \tilde \sigma _{12} A_{1y} A_{2y} }  \\
   0  \\
\end{pmatrix}_{z = 0^ +  },
\end{eqnarray}
where, for each component, we have set $\tilde \sigma  = \sqrt {\mu _0 /\varepsilon _0 } \sigma$, where $\varepsilon _0$ and $\mu _0$ indicate the dielectric permittivity and magnetic permeability of vacuum, respectively. After inserting Eqs. (\ref{boundaryfields}) along with $E_1^{\left( i \right)} = E_2^{\left( i \right)} =0$ into Eqs. (\ref{matching}) we obtain
\begin{eqnarray} \label{outputfields}
 V_{1x}^{\left( t \right)} E_1^{\left( t \right)}  - V_{1x}^{\left( r \right)} E_1^{\left( r \right)}  &=& \tilde \sigma _1 V_{1y}^{\left( t \right)} E_1^{\left( t \right)}  + \tilde \sigma _{23} V_{2y}^{\left( t \right)*} V_{3y}^{\left( t \right)} E_2^{\left( t \right)*} E_3^{\left( t \right)} , \nonumber \\
 V_{1y}^{\left( t \right)} E_1^{\left( t \right)}  - V_{1y}^{\left( r \right)} E_1^{\left( r \right)}  &=& 0, \nonumber \\
 V_{2x}^{\left( t \right)} E_2^{\left( t \right)}  - V_{2x}^{\left( r \right)} E_2^{\left( r \right)}  &=& \tilde \sigma _2 V_{2y}^{\left( t \right)} E_2^{\left( t \right)}  + \tilde \sigma _{13} V_{1y}^{\left( t \right)*} V_{3y}^{\left( t \right)} E_1^{\left( t \right)*} E_3^{\left( t \right)} , \nonumber \\
 V_{2y}^{\left( t \right)} E_2^{\left( t \right)}  - V_{2y}^{\left( r \right)} E_2^{\left( r \right)}  &=& 0, \nonumber \\
 V_{3x}^{\left( t \right)} E_3^{\left( t \right)}  - V_{3x}^{\left( i \right)} E_3^{\left( i \right)}  - V_{3x}^{\left( r \right)} E_3^{\left( r \right)}  &=& \tilde \sigma _3 V_{3y}^{\left( t \right)} E_3^{\left( t \right)}  + \tilde \sigma _{12} V_{1y}^{\left( t \right)} V_{2y}^{\left( t \right)} E_1^{\left( t \right)} E_2^{\left( t \right)} , \nonumber \\
 V_{3y}^{\left( t \right)} E_3^{\left( t \right)}  - V_{3y}^{\left( i \right)} E_3^{\left( i \right)}  - V_{3y}^{\left( r \right)} E_3^{\left( r \right)}  &=& 0,
\end{eqnarray}
which are six equations for the six unknown amplitudes $E_n^{\left( t \right)}$, $E_n^{\left( r \right)}$ of the signal, idler and pump fields transmitted and reflected by the cavity illuminated by incident pump field of amplitude $E_3^{\left( i \right)}$. The second, fourth and sixth of Eqs. (\ref{outputfields}) can be written as
\begin{eqnarray} \label{reflectedfields}
 E_1^{\left( r \right)}  &=& \frac{{V_{1y}^{\left( t \right)} }}{{V_{1y}^{\left( r \right)} }}E_1^{\left( t \right)}, \nonumber \\
 E_2^{\left( r \right)}  &=& \frac{{V_{2y}^{\left( t \right)} }}{{V_{2y}^{\left( r \right)} }}E_2^{\left( t \right)}, \nonumber \\
 E_3^{\left( r \right)}  &=& \frac{{V_{3y}^{\left( t \right)} }}{{V_{3y}^{\left( r \right)} }}E_3^{\left( t \right)}  - \frac{{V_{3y}^{\left( i \right)} }}{{V_{3y}^{\left( r \right)} }}E_3^{\left( i \right)},
\end{eqnarray}
showing that the reflected fields can be evaluated once the transmitted fields are known. Substituting the reflected fields from Eqs. (\ref{reflectedfields}) into Eqs. (\ref{outputfields}) and using Eqs. (\ref{amplitudesV}) we eventually get
\begin{eqnarray} \label{transmittedfields}
 \Delta _1 Q_1  + \tilde \sigma _{23} Q_2^* Q_3  &=& 0, \nonumber \\
 \Delta _2 Q_2  + \tilde \sigma _{13} Q_1^* Q_3  &=& 0, \nonumber \\
 \Delta _3 Q_3  + \tilde \sigma _{12} Q_1 Q_2  &=& P_3,
\end{eqnarray}
where
\begin{eqnarray} \label{QPD}
 Q_n  &=& \left( {\frac{{r_n^{\left( R \right)} e^{iq_n L}  + e^{ - iq_n L} }}{{t_n^{\left( R \right)} }}} \right)E_n^{\left( t \right)}, \nonumber \\
 P_3  &=& \left( {\frac{{2t_3^{\left( R \right)} \cos \theta }}{{r_3^{\left( R \right)}  + 1}}} \right)E_3^{\left( i \right)}, \nonumber \\
 \Delta _n  &=& \tilde \sigma _n  - \frac{c q_n }{\omega_n}\left( {\frac{{r_n^{\left( R \right)}  - 1}}{{r_n^{\left( R \right)}  + 1}} + \frac{{r_n^{\left( R \right)} e^{iq_n L}  - e^{ - iq_n L} }}{{r_n^{\left( R \right)} e^{iq_n L}  + e^{ - iq_n L} }}} \right).
\end{eqnarray}
Equations (\ref{transmittedfields}) are the basic equations for the output fields. Note that Eqs. (\ref{transmittedfields}) have been derived without resorting to any electromagnetic approximation commonly used in cavity nonlinear optics (e.g. slowly varying amplitude approximation, decoupled approximation for counter-propagating waves, etc.) and this is a consequence of the fact the overall nonlinear response of the MX$_2$ monolayer is confined to a single plane.

\section{Doubly Resonant Parametric Oscillation conditions}
Equations (\ref{transmittedfields}) provide the amplitudes $Q_n$ (proportional to the amplitudes of the transmitted signal, idler and pump fields) for a given amplitude $P_3$ (proportional to the amplitude of the incident pump field). Note that they always admit the solution
\begin{equation}
Q_1 = Q_2 = 0, \quad Q_3  = \frac{P_3}{\Delta_3},
\end{equation}
which describes the linear response of the cavity to the pump field without parametric oscillations (POs) in turn characterized by $Q_1 \neq 0$ and $Q_2 \neq 0$. On the other hand, the first and the complex conjugate of the second of Eqs. (\ref{transmittedfields}) are a linear system for $Q_1$ and $Q_2^*$ and it admits nontrivial solutions only if its determinant vanishes (see Section IV below) or
\begin{equation} \label{POsufficient}
\left| {Q_3 } \right|^2  = \frac{{\Delta _1 \Delta _2^* }}{{\tilde \sigma _{23} \tilde \sigma _{13}^* }}.
\end{equation}
This condition entails the occurrence of POs since if it is fulfilled, Eqs. (\ref{transmittedfields}) have solutions with $Q_1 \neq 0$ and $Q_2 \neq 0$ which are stable whereas the linear one ($Q_1 = Q_2 = 0$) becomes unstable. Since the right hand side of Eq. (\ref{POsufficient}) is generally a complex number, it is evident that POs can occur {\it only if} such complex number is real and positive or
\begin{equation} \label{POnecessary}
 \frac{{\Delta _1 \Delta _2^* }}{{\tilde \sigma _{23} \tilde \sigma _{13}^* }}=
\left| \frac{{\Delta _1 \Delta _2^* }}{{\tilde \sigma _{23} \tilde \sigma _{13}^* }} \right|,
\end{equation}
which, due to Eqs. (\ref{frequencies}) and (\ref{QPD}), is a constraint joining the cavity length $L$, the note-beat frequency $\Delta \omega$, and the incident angle $\theta$. Geometrically, Eq. (\ref{POnecessary}) represents a surface $\Sigma$ of the three-dimensional cavity state space $(L,\Delta \omega,\theta)$, and its typical slices ($\theta = 0$ or $L = L_0$) are illustrated in Figs.3a1, 3b1 and 3c1 of the main paper (green curves). At each point of the surface $\Sigma$, PO ignites if $|Q_3|$ is sufficiently large to fulfill Eq. (\ref{POsufficient}). Therefore, considering an experiment where the incident pump $|P_3|$ is gradually increased starting from the linear regime where $Q_1=Q_2=0$, the PO threshold  is obtained by inserting the linear solution $Q_3  = \frac{P_3}{\Delta _3}$ into Eq. (\ref{POsufficient}), thus obtaining
\begin{equation} \label{threshold}
\left( \left| {P_3 } \right|^2 \right)_{th} = \frac{{\Delta _1 \Delta _2^* }}{{\tilde \sigma _{23} \tilde \sigma _{13}^* }}\left| {\Delta _3 } \right|^2,
\end{equation}
which, through the second of Eqs. (\ref{QPD}) and the relation $I_3^{(i)} = \frac{1}{2}\sqrt{\frac{\epsilon_0}{\mu 0}} \left| E_3^{\left( i \right)} \right|^2$, entails the intensity threshold for the incident pump. Note that the denominator of the right hand side of Eq. (\ref{threshold}) contains the nonlinear conductivities $\tilde \sigma _{23}$ and $\tilde \sigma _{13}$ whose moduli are so small to generally yield exceedingly large and unfeasible intensity thresholds. A viable way for observing POs thus necessitates the identification of the points $(L,\Delta \omega,\theta)$ of the surface $\Sigma$ for which $|\Delta_1|$ and $|\Delta _2|$ are very close to zero. An inspection of the third of Eqs. (\ref{QPD}) reveals that $|\Delta _n|$ can not be small if $\left|r_n^{\left( R \right)}\right|$ is not close to $1$. Therefore, by choosing Bragg mirrors with high reflectivities, the third of Eqs. (\ref{QPD}) can be expanded up to the first order in the parameter $1 - \left| {r_n^{\left( R \right)} } \right| \ll 1$ thus getting
\begin{equation}
\Delta _n  = \left[ {\tilde \sigma _n  - i\left( {\frac{{2 c q_n }}{{\omega_n }}} \right)\frac{{\sin \Gamma _n }}{{\cos \Gamma _n  + \cos \left( {q_n L} \right)}}} \right] + \left\{ {\left( {\frac{{2 c q_n }}{{\omega_n }}} \right)\frac{{1 + \left( {\cos \Gamma _n  + 2i\sin \Gamma _n } \right)\cos \left( {q_n L} \right)}}{{\left[ {\cos \Gamma _n  + \cos \left( {q_n L} \right)} \right]^2 }}} \right\}\left( {1 - \left| {r_n^{\left( R \right)} } \right|} \right),
\end{equation}
where $\Gamma_n  = q_n L + \arg  {r_n^{\left( R \right)} } $. The minima of $|\Delta_n|$ are easily seen to occur for $\Gamma_n = m \pi$ (where $m$ is any integer) which is exactly the cavity resonance condition for the frequency $\omega_n$ \cite{Kavok} and where, up to the first order of $1 - \left| {r_n^{\left( R \right)} } \right|$,
\begin{equation} \label{Delta_res}
\Delta _n  = \tilde \sigma _n  + \left[ {\left( {\frac{{2 c q_n }}{{\omega_n }}} \right)\frac{1}{{1 + \cos \left( {\arg r_n^{\left( R \right)} } \right)}}} \right] \left( {1 - \left| {r_n^{\left( R \right)} } \right|} \right).
\end{equation}
Therefore, as for standard POs based on bulk nonlinear media, the pump intensity threshold is here minimum when one or more of the three fields meet the cavity resonant condition. Designing a Bragg mirror with high reflectivity for both signal and idler fields is relatively simple (see the Section III below) and therefore in this paper we consider only doubly resonant (DR) states where the signal resonance (SR) and idler resonance (IR) conditions
\begin{eqnarray} \label{s1_s2}
 q_1 L + \arg r_1^{\left( R \right)}  &=& m_1 \pi, \nonumber \\
 q_2 L + \arg r_2^{\left( R \right)}  &=& m_2 \pi,
\end{eqnarray}
are both achieved whereas the pump is non-resonant. Such two equations represents two surfaces $\Sigma_1$ and $\Sigma_2$ of the space   $(L,\Delta \omega,\theta)$ whose typical slices are reported in Figs.3a1, 3b1 and 3c1 of the main text (black and red lines respectively) and whose intersection describes the DR cavity states where $|\Delta_1 \Delta_2|$ is minimum. The (nontrivial) intersection among the three surfaces $\Sigma$, $\Sigma_1$ and $\Sigma_2$ is the set of the DRPO cavity states with feasible intensity threshold. Note that if $\Sigma_1$ and $\Sigma_2$ intersect at a specific $(L,0,\theta)$ point this point also belongs to the surface $\Sigma$ since for $\Delta \omega = 0$ Eqs. (\ref{POnecessary}) is trivially satisfied since evidently $\omega_1 = \omega_2$, $\Delta_1 = \Delta_2$ and $\sigma_{23} = \sigma_{13}$. In other words a degenerate ($\omega_1 = \omega_2$) DR state always supports a PO which we refer to as a degenerate DRPO. In addition, note that if $|{\rm Re} \: \tilde \sigma _n| \ll |{\rm Im} \: \tilde \sigma _n|$ and $|{\rm Re} \: \tilde \sigma _{nm}| \ll |{\rm Im} \: \tilde \sigma _{nm}|$, if both Eqs. (\ref{s1_s2}) are satisfied with $\Delta \omega \neq 0$, Eq. (\ref{Delta_res}) implies that $\left| {\rm Im} \left( \frac{{\Delta _1 \Delta _2^* }}{{\tilde \sigma _{23} \tilde \sigma _{13}^* }} \right) \right| \ll \left|{\rm Re} \left( \frac{{\Delta _1 \Delta _2^* }}{{\tilde \sigma _{23} \tilde \sigma _{13}^* }} \right) \right|$ so that, remarkably, if both linear and nonlinear absorption are small the non-degenerate DR states are always very close to PO states. As a consequence the non-degenerate DRPOs with feasible intensity threshold are associated to those points of the surface $\Sigma$ which are as close as possible to points of the intersection between the surface $\Sigma_1$ and $\Sigma_2$. Both degenerate and non-degenerate DRPO states are labelled with a dashed disk in Figs.3a1, 3b1 and 3c1 of the main text.

\begin{figure}[b]
\center
\includegraphics*[width=0.8\textwidth]{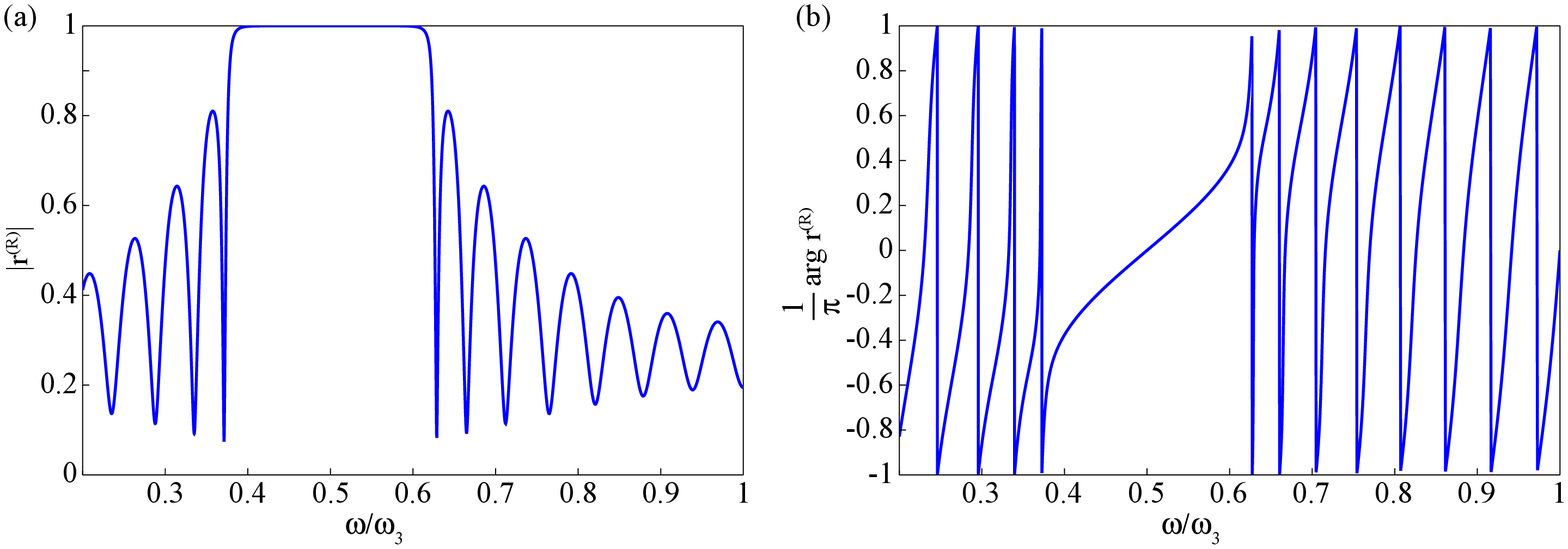}
\caption{Absolute value (a) and argument (b) of the Bragg Mirror complex reflectivity $r^{(R)}$ for normal incidence $\theta =0$}
\label{S2}
\end{figure}

\section{Bragg mirror design}
The Bragg mirror is a periodic structure composed of $N$ bi-layers whose dielectric materials have refractive indexes $n^{(a)}$ and $n^{(b)} $ and thicknesses $a$ and $b$. If the layers' thicknesses are chosen to satisfy the Bragg interference condition
\begin{equation}
a n^{(a)} = b n^{(b)} = \frac{\pi c}{2 \bar \omega},
\end{equation}
the mirror has (for normal incidence $\theta = 0$), a spectral stop-band centered at $\bar \omega$ whose width is proportional to the refractive index contrast $|n^{(a)} - n^{(b)}|$ \cite{Kavok}. Within the stop-band the mirror reflectivity is very large, the larger $N$ the closer $|r^{(R)} (\omega)|$ to 1. As explained above in Section II, in order for one of the three fields (pump, signal and idler) to be resonant, it is necessary a very large reflectivity of the Bragg mirror at the field angular frequency, or in other words the frequency $\omega_n$ has to lie within the mirror stop-band. As noted above, the degenerate DR states with $\Delta \omega = 0$ (where, from Eqs. (\ref{frequencies}), $\omega_1= \omega_2 =\omega_3/2$) rigorously supports POs so that it is convenient to set the center of the mirror stop-bad at $\bar \omega = \omega_3/2$. Due to the refractive index contrast $|n^{(a)} -n^{(b)}|$, this condition assures that both signal and idler fields experience very large mirror reflectivity in a range of $\Delta \omega$ and can accordingly be resonant at the same time. On the other hand the pump frequency $\omega_3$ is twice the central mirror frequency $\bar \omega$ and requiring also the pump to resonate would require very large refractive contrast. To avoid this difficulty we have chosen to leave the pump out of resonance.

In the analysis reported in Fig.3 of the main text, we have set as pump wavelength $\lambda_3 = 780 \: {\rm nm}$. For the Bragg Mirror we have chosen the refractive indexes $n_a = 1.2$ and $n_b = 2.5$ so that, in order to have the center of the stop-band at $\bar \omega = \omega_3/2$ we have chosen the thicknesses $a = \lambda_3/(2 n^{(a)}) = 325 \: {\rm nm}$ and $b = \lambda_3/(2 n^{(b)}) = 156 \: {\rm nm}$. We have also set $N=8$ for dealing with an efficient, feasible and compact Bragg mirror of length $d = N(a+b) =  3848 \: {\rm nm}$. Using the transfer matrix approach, the complex reflectivity $r^{(R)}$ (for normal incidence $\theta=0$) of the Bragg mirror which has vacuum and the dielectric at its left and right sides, respectively, is easily evaluated and we plot its absolute value and argument in panel (a) and (b), respectively, of Fig.\ref{S2}. Accordingly, the Bragg mirror stop-band is centered at $\omega_3/2$ and its spectral width is $\simeq 0.22$ $\omega_3$. As a consequence, if $\omega_1$ and $\omega_2$ lie within this stop-band, signal and idler waves can resonate simultaneously since $r_1^{(B)} = r^{(B)} (\omega_1)$ and $r_2^{(B)} = r^{(B)} (\omega_2)$ have moduli very close to $1$. The mirror stop-band width therefore yields the note-beat frequency range $0 \le \Delta \omega < 0.22 \omega$, which is the one considered in the analysis reported in Fig.3 of the main text. Note that $\omega_3$ lies outside the mirror stop-band and thus the pump field does not resonate.

\section{Output Fields}
In order to evaluate the PO output fields $E_n^{\left( t \right)}$, Eqs. (\ref{transmittedfields}) have to be solved for a given incident pump field $E_3^{\left( i\right)}$. POs are characterized by $Q_1 \neq 0$ and $Q_2 \neq 0$ for a given $Q_3 \neq 0$. First note that Eqs. (\ref{transmittedfields}) are left invariant by the gauge transformation
\begin{eqnarray}
 Q_1  &\to& Q_1 e^{i \theta }, \nonumber  \\
 Q_2  &\to& Q_2 e^{ - i \theta },
\end{eqnarray}
and this implies that for a given $P_3$ there are infinite pairs $(Q_1,Q_2)$, all with the same $\Psi = \arg Q_1 + \arg Q_2$. In other words, the phase difference $\Phi = \arg Q_1 - \arg Q_2$ is not set by the pump field $P_3$. Evidently, such symmetry is spontaneously broken in actual experiments where a single pair $(Q_1,Q_2)$ (i.e. a single value of $\Phi$) is selected by the specific way chosen to trigger POs.

In the case of POs, the first and the complex conjugate of the second of Eqs. (\ref{transmittedfields}) can be casted as
\begin{eqnarray} \label{temp}
\frac{{Q_1 }}{{Q_2^* }} = - \frac{{\tilde \sigma _{23} Q_3}}{{\Delta _1 }}, \nonumber \\
\frac{{Q_1 }}{{Q_2^* }} =  - \frac{{\Delta _2^* }}{{\tilde \sigma _{13}^* Q_3^* }},
\end{eqnarray}
whose consistency requires their right and left hand sides to coincide or
\begin{equation} \label{POsufficient2}
\left| {Q_3 } \right|^2  = \frac{{\Delta _1 \Delta _2^* }}{{\tilde \sigma _{23} \tilde \sigma _{13}^* }}
\end{equation}
which is the PO oscillation condition of Section II. [see Eq. (\ref{POsufficient})]. The right hand side of Eq. (\ref{POsufficient2}) is a positive real number if and only if the complex numbers $\Delta _1 /\tilde \sigma _{23}$ and $\Delta _2 /\tilde \sigma _{13}$ have the same argument $\varphi$ or
\begin{eqnarray} \label{POnecessary2}
 \Delta _1  &=& \tilde \sigma _{23} \left| {\frac{{\Delta _1 }}{{\tilde \sigma _{23} }}} \right|e^{i\varphi }, \nonumber \\
 \Delta _2  &=& \tilde \sigma _{13} \left| {\frac{{\Delta _2 }}{{\tilde \sigma _{13} }}} \right|e^{i\varphi },
\end{eqnarray}
which are equivalent to Eq. (\ref{POnecessary}) of Section II so that, considering only those states for which Eqs. (\ref{POnecessary2}) are satisfied, Eqs. (\ref{transmittedfields}) yield
\begin{eqnarray} \label{transmittedfields2}
 \left| {\frac{{\Delta _1 }}{{\tilde \sigma _{23} }}} \right|e^{i\varphi } Q_1  + Q_2^* Q_3  &=& 0, \nonumber \\
 \left| {\frac{{\Delta _2 }}{{\tilde \sigma _{13} }}} \right|e^{i\varphi } Q_2  + Q_1^* Q_3  &=& 0, \nonumber \\
 \Delta _3 Q_3  + \tilde \sigma _{12} Q_1 Q_2  &=& P_3.
\end{eqnarray}
To solve this equation, after noting that Eqs. (\ref{temp}) require that $\left|\frac {Q_1 } {Q_2 } \right|^2  = \left|\frac{ \Delta _2{\tilde \sigma _{23} }}{{\Delta _1 \tilde \sigma _{13} }} \right|$ and exploiting the above discussed gauge symmetry, we set
\begin{eqnarray} \label{change}
Q_1  &=& \sqrt {\left| {\frac{{\Delta _2 }}{{\tilde \sigma _{13} }}} \right|} \left| Q \right|e^{i\frac{1}{2}\left( {\Psi  + \Phi } \right)}, \nonumber  \\
Q_2  &=& \sqrt {\left| {\frac{{\Delta _1 }}{{\tilde \sigma _{23} }}} \right|} \left| Q \right|e^{i\frac{1}{2}\left( {\Psi  - \Phi } \right)}, \end{eqnarray}
where we have used the symbol $|Q|$ to stress than this quantity is real. Inserting Eqs. (\ref{change}) into Eqs. (\ref{transmittedfields2}) yield
\begin{eqnarray} \label{transmittedfields3}
 Q_3  &=& - \sqrt {\left| {\frac{{\Delta _1 \Delta _2 }}{{\tilde \sigma _{23} \tilde \sigma _{13} }}} \right|} e^{i\left( {\Psi  + \varphi } \right)} , \nonumber \\
\left| Q \right|^2  +  \frac{{\Delta _3 }}{{\tilde \sigma _{12} }} \sqrt {\left| {\frac{{\tilde \sigma _{23} \tilde \sigma _{13} }}{{\Delta _1 \Delta _2 }}} \right|} e^{ - i\Psi } Q_3  &=& \sqrt {\left| {\frac{{\tilde \sigma _{23} \tilde \sigma _{13} }}{{\Delta _1 \Delta _2 }}} \right|} \frac{{P_3 }}{{\tilde \sigma _{12} }}e^{ - i\Psi }.
\end{eqnarray}
Note that the first two of Eqs. (\ref{transmittedfields2}) both reduces to the first of Eqs. (\ref{transmittedfields3}) through the change of variables given by Eqs. (\ref{change}) and this is a consequence of the necessary Eq. (\ref{POnecessary2}). Substituting $Q_3$ from the first of Eqs. (\ref{transmittedfields3}) into the second we get
\begin{equation} \label{single}
\left| Q \right|^2  - \frac{{\Delta _3 }}{{\tilde \sigma _{12} }}e^{i\varphi }  =
\sqrt {\left| \frac {{\tilde \sigma _{23} \tilde \sigma _{13} }} {{\Delta _1 \Delta _2 }} \right|}
\frac{{P_3 }}{{\tilde \sigma _{12} }}
e^{ - i\Psi },
\end{equation}
which is a single complex equation for $|Q|$ and $\Psi$. After equating the square-moduli of the left and right sides of this equation we get
\begin{equation}
\left| Q \right|^4  - 2{\mathop{\rm Re}\nolimits} \left( {\frac{{\Delta _3 }}{{\tilde \sigma _{12} }}e^{i\varphi } } \right)\left| Q \right|^2  + \left( \left| {\frac{{\Delta _3 }}{{\tilde \sigma _{12} }}e^{i\varphi } } \right|^2  -
\left| \frac {{\tilde \sigma _{23} \tilde \sigma _{13} }} {{\Delta _1 \Delta _2 }} \right|
\left| {\frac{{P_3 }}{{\tilde \sigma _{12} }}} \right|^2 \right) = 0,
\end{equation}
which is a biquadratic equation for $\left| Q \right|$ whose solutions are
\begin{equation} \label{Q}
\left| Q \right| = \sqrt {{\mathop{\rm Re}\nolimits} \left( {\frac{{\Delta _3 }}{{\tilde \sigma _{12} }}e^{i\varphi } } \right) +\xi \sqrt {\left| \frac {{\tilde \sigma _{23} \tilde \sigma _{13} }} {{\Delta _1 \Delta _2 }} \right|
\left| {\frac{{P_3 }}{{\tilde \sigma _{12} }}} \right|^2
 - {\mathop{\rm Im}\nolimits}^2 \left( {\frac{{\Delta _3 }}{{\tilde \sigma _{12} }}e^{i\varphi } } \right) } },
\end{equation}
where $\xi = \pm 1$. Note that, due to the $\xi$ factor, generally there are two $|Q|$ corresponding to a $|P_3|$ and hence bistable POs can in principle occur. In addition, it is fundamental stressing that $|Q|$ is real and hence Eq. (\ref{Q}) provides its value only if the arguments of the square roots are positive. Before discussing the range of $|P_3|$ where this is the case (see below), we assume $|Q|$ real and we deduce the output field amplitudes. Equation (\ref{single}) yields
\begin{equation}
e^{i\Psi }  = \sqrt {\left| {\frac{{\tilde \sigma _{23} \tilde \sigma _{13} }}{{\Delta _1 \Delta _2 }}} \right|} \frac{\displaystyle {P_3 }}{\displaystyle {\tilde \sigma _{12} \left( {\left| Q \right|^2  - \frac{{\Delta _3 }}{{\tilde \sigma _{12} }}e^{i\varphi } } \right)}},
\end{equation}
which is consistent since the modulus of its right hand side, due to Eq. (\ref{Q}), is equal to $1$. Hence Eqs. (\ref{change}) and the first of Eqs. (\ref{transmittedfields3}) eventually yield
\begin{eqnarray} \label{Q1Q2Q3}
 Q_1  &=& \zeta \left| {\frac{\displaystyle {\Delta _2 \tilde \sigma _{23} }}{\displaystyle {\Delta _1 \tilde \sigma _{13} }}} \right|^{1/4} \sqrt {\frac{\displaystyle {P_3 }}{\displaystyle {\tilde \sigma _{12} \left( {\left| Q \right|^2  - \frac{{\Delta _3 }}{{\tilde \sigma _{12} }}e^{i\varphi } } \right)}}} e^{i\frac{\Phi }{2}} \left| Q \right|, \nonumber \\
 Q_2  &=& \zeta \left| {\frac{\displaystyle {\Delta _1 \tilde \sigma _{13} }}{\displaystyle {\Delta _2 \tilde \sigma _{23} }}} \right|^{1/4} \sqrt {\frac{\displaystyle {P_3 }}{\displaystyle {\tilde \sigma _{12} \left( {\left| Q \right|^2  - \frac{{\Delta _3 }}{{\tilde \sigma _{12} }}e^{i\varphi } } \right)}}} e^{ - i\frac{\Phi }{2}} \left| Q \right|, \nonumber \\
 Q_3  &=& \frac{\displaystyle {P_3 }}{\displaystyle {\tilde \sigma _{12} \left( {\left| Q \right|^2  - \frac{{\Delta _3 }}{{\tilde \sigma _{12} }}e^{i\varphi } } \right)}}e^{i\left( {\varphi  + \pi } \right)},
\end{eqnarray}
where $\zeta = \pm 1$ and the principal branch is assumed for all the complex square-roots.

Note that the output fields of Eqs. (\ref{Q1Q2Q3}) satisfy Eq. (\ref{POsufficient2}) so that they describe all the possible cavity POs whenever they exist or, in other words, whenever $|Q|$ of Eq. (\ref{Q}) is a positive real number. Such requirement evidently sets a range for the input pump intensity $|P_3|^2$ and there are four different cases corresponding to the two values of $\xi$ and of the two signs of
${\mathop{\rm Re}\nolimits} \left( {\frac{{\Delta _3 }}{{\tilde \sigma _{12} }}e^{i\varphi } } \right)$. The results of this analysis are reported in the following table.
\begin{equation*}
\begin{array}{c | c | c }
    & {\mathop{\rm Re}\nolimits} \left( {\frac{{\Delta _3 }}{{\tilde \sigma _{12} }}e^{i\varphi } } \right)>0  & {\mathop{\rm Re}\nolimits} \left( {\frac{{\Delta _3 }}{{\tilde \sigma _{12} }}e^{i\varphi } } \right) <0 \\
\hline
    & & \\
   \xi=1  &
\left| {P_3 } \right|^2  > \left( \left| {P_3 } \right|^2 \right)_{-} & \left| {P_3 } \right|^2  > \left( \left| {P_3 } \right|^2 \right)_{th} \\
    & &  \\
    \hline
    & &  \\
   \xi=-1 & \left( \left| {P_3 } \right|^2 \right)_{-} < \left| {P_3 } \right|^2  < \left( \left| {P_3 } \right|^2 \right)_{th} &
{\rm no} \: |P_3|^2 \\
    & &  \\
    \hline
\end{array}
\end{equation*}
Here we have set
\begin{equation}
\left( \left| {P_3 } \right|^2 \right)_{-} = \left| {\frac{{\Delta _1 \Delta _2 }}{{\tilde \sigma _{23} \tilde \sigma _{13} }}} \right|{\mathop{\rm Im}\nolimits} ^2 \left( {\Delta _3 \frac{{\left| {\tilde \sigma _{12} } \right|}}{{\tilde \sigma _{12} }}e^{i\varphi } } \right) <
{\left| {\frac{{\Delta _1 \Delta _2 }}{{\tilde \sigma _{23} \tilde \sigma _{13} }}} \right|\left| {\Delta _3 } \right|^2 } =
\left( \left| {P_3 } \right|^2 \right)_{th}.
\end{equation}

Therefore at each state where PO can occur (i.e. at a each point of the surface $\Sigma$) the scenario is the following one. If ${\mathop{\rm Re}\nolimits} \left( {\frac{{\Delta _3 }}{{\tilde \sigma _{12} }}e^{i\varphi } } \right) < 0$, there is only one PO (with $\xi =1$) that effectively starts when Eq. (\ref{threshold}) is satisfied, thus confirming the analysis of Section III. On the other hand, if ${\mathop{\rm Re}\nolimits} \left( {\frac{{\Delta _3 }}{{\tilde \sigma _{12} }}e^{i\varphi } } \right) > 0$ the scenario changes qualitatively since in this case there are two allowed POs (with $\xi=1$ and $\xi=-1$) when  $\left( \left| {P_3 } \right|^2 \right)_{-} < \left| {P_3 } \right|^2  < \left( \left| {P_3 } \right|^2 \right)_{th}$  and a single PO (with $\xi=1$) when Eq. (\ref{threshold}) is satisfied. As a consequence in this case POs also exist {\it below} the threshold.

In order to grasp the reason why the sub-threshold POs have not been entailed in Section III, note that in the case ${\mathop{\rm Re}\nolimits} \left( {\frac{{\Delta _3 }}{{\tilde \sigma _{12} }}e^{i\varphi } } \right) > 0$, if $\left| {P_3 } \right|^2 = \left( \left| {P_3 } \right|^2 \right)_{-}$, Eq. (\ref{Q}) implies that $\left| Q \right| = \sqrt {{\mathop{\rm Re}\nolimits} \left( {\frac{{\Delta _3 }}{{\tilde \sigma _{12} }}e^{i\varphi } } \right)}  \ne 0$ so that $Q_1 \neq 0$ and $Q_2 \neq 0$. In other words in this situation the signal and idler fields do not vanish {\it at the threshold} and accordingly this case is ruled out from the reasoning of Section III where the threshold has been obtained for PO oscillation starting from the linear regime. In a realistic experiment PO is switched on starting from the linear regime, with the intensity threshold given by Eq. (\ref{threshold}). However, once PO ignites, by changing the pump intensity, incidence angle $\theta$, or the cavity length, we argue that one can in principle access sub-threshold PO states. In every design considered in the main paper we have focused on the case ${\mathop{\rm Re}\nolimits} \left( {\frac{{\Delta _3 }}{{\tilde \sigma _{12} }}e^{i\varphi } } \right) < 0$, where sub-threshold PO does not occur.

\begin{figure}[t]
\center
\includegraphics[width=0.8\textwidth]{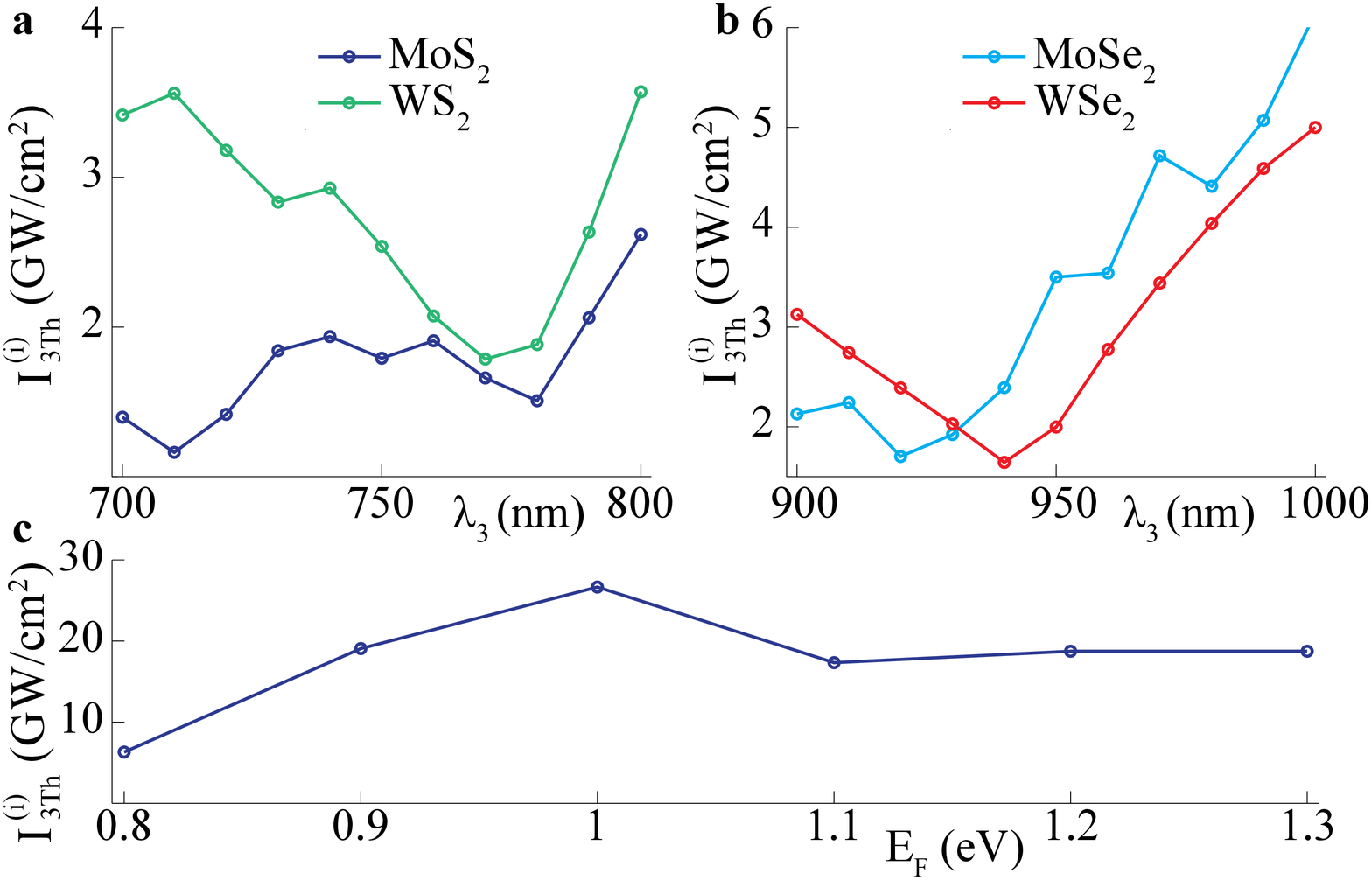}
\caption{{\bf Parametric Oscillation Thresholds}. {\bf a,b} Pump wavelength $\lambda_3$ dependence of the pump intensity thresholds for parametric oscillators embedding MoS$_2$, WS$_2$ ({\bf a}) and MoSe$_2$, WSe$_2$ ({\bf b}). {\bf c} Pump intensity threshold of a parametric oscillator embedding MoS$_2$ as a function of the Fermi level. The thresholds are associated to non-degenerate ($\Delta \omega = 0$) POs and they have been calculated for cavities whose length $L$ is equal to the pump wavelength $\lambda_3$.}
\label{figS3}
\end{figure}

\section{Pump intensity thresholds}
In Fig. \ref{figS3} we compare the calculated pump intensity thresholds versus the pump wavelength $\lambda_3$ for parametric oscillators embedding MoS$_2$, WS$_2$ (Fig. \ref{figS3}a) and MoSe$_2$, WSe$_2$ (Fig. \ref{figS3}b). Note that, while the minimal pump intensity threshold occurs at $\lambda_3 \approx 780$ nm for MoS$_2$ and WS$_2$, it shifts to $\lambda_3 \approx 940$ nm for MoSe$_2$ and WSe$_2$. None of the ML-TMDs examined enables feasible PO with low pump intensity threshold at optical frequencies owing to the enlarged absorption in this frequency range, which is the main responsible for oscillation quenching. In addition, at optical frequencies such materials exhibit exciton resonances \cite{UBS2014} (not taken into account in our theoretical approach) that are also detrimental for POs owing to the enhanced absorption they are accompanied with. In Fig. \ref{figS3}c we plot the pump intensity threshold as a function of the Fermi level of MoS$_2$, showing that it can be increased efficiently. Thus, the external gate voltage quenches POs when the optical pump is fixed and fast modulation of the output signal and idler fields can be achieved with novel parametric oscillators embedding ML-TMDs.

\end{document}